\documentclass[preprint,prc,showpacs,preprintnumbers]{revtex4}
\usepackage{graphicx,latexsym,amssymb,amsmath,color,multirow,mathrsfs,booktabs,ifpdf}
\usepackage{dcolumn}
\usepackage{bm}
\usepackage{longtable}
\begin{document}
\title{Mean-field study of hot $\beta$-stable protoneutron 
star matter: Impact of the symmetry energy and nucleon effective mass}
\author{Ngo Hai Tan$^{1,2}$}
\author{Doan Thi Loan$^1$}
\author{Dao T. Khoa$^1$}\email{khoa@vinatom.gov.vn}
\author{Jerome Margueron$^2$}
\affiliation{$^1$ Institute for Nuclear Science and Technology, VINATOM \\ 
179 Hoang Quoc Viet, Cau Giay, Hanoi, Vietnam. \\
$^2$ Institut de Physique Nucl\'eaire de Lyon, IN2P3-CNRS \\
4 Rue Enrico Fermi, 69622 Villeurbanne Cedex, France.}
\begin{abstract}{A consistent Hartree-Fock study of the equation of state (EOS) 
of asymmetric nuclear matter at finite temperature has been performed using 
realistic choices of the effective, density dependent nucleon-nucleon (NN) 
interaction, which were successfully used in different nuclear structure and 
reaction studies. Given the importance of the nuclear symmetry energy in the 
neutron star formation, EOS's associated with different behaviors of the 
symmetry energy were used to study hot asymmetric nuclear matter. The slope 
of the symmetry energy and nucleon effective mass with increasing baryon 
density was found to affect the thermal properties of nuclear matter 
significantly. Different density dependent NN interactions were further used 
to study the EOS of hot protoneutron star (PNS) matter of the $npe\mu\nu$ 
composition in $\beta$-equilibrium. The hydrostatic configurations of PNS in 
terms of the maximal gravitational mass $M_{\rm max}$ and radius, central density, 
pressure and temperature at the total entropy per baryon $S/A= 1,2$ and 4 have 
been determined in both the neutrino-free and neutrino-trapped scenarios. 
The obtained results show consistently a strong impact of the symmetry energy 
and nucleon effective mass on thermal properties and composition of hot PNS 
matter. $M_{\rm max}$ values obtained for the (neutrino-free) $\beta$-stable 
PNS at $S/A=4$ were used to assess time $t_{\rm BH}$ of the collapse 
of 40 $M_\odot$ protoneutron progenitor to black hole, based on a correlation 
between $t_{\rm BH}$ and $M_{\rm max}$ found from the hydrodynamic simulation 
by Hempel {\it et al.}.}
\end{abstract}
\date{\today}
\maketitle

\section{Introduction}
\label{intro} 
A realistic equation of state (EOS) of asymmetric nuclear matter 
(NM) is the most vital input for the study of the neutron star formation
\cite{Bet90,Bur86,Lat04,Lat07,Hae07,Shapiro}, and the determination of the EOS has 
been, therefore, the focus of many nuclear structure and reaction studies
involving exotic nuclei lying close to the driplines. The key quantity
to distinguish the different nuclear EOS's is the nuclear mean-field potential 
that can be obtained from a consistent mean-field study, like the relativistic
mean-field (RMF) \cite{Shen} or Hartree-Fock (HF) \cite{Loa11} calculations 
of NM using the realistic interaction between nucleons in the high-density 
nuclear medium. As such, the different nuclear EOS's have been used in the
hydrodynamic simulation of core-collapse supernovae \cite{Hem12,Ste13} to 
study the hot protoneutron star (PNS) formed in the aftermath of supernovae 
\cite{Pra97,Bur10,Li10} and its collapse to black hole or evolution towards 
cold neutron star. The latter is an extensive research object of numerous nuclear 
many-body studies (see, e.g., Refs.~\cite{Dou01,Kl06,Bal07,Ba08,Tha09,Loa11}).

The microscopic many-body studies, like the Brueckner-Hartree-Fock (BHF) 
calculations \cite{Bur10,Bal07,Tar13}, have shown the important role played 
by the Pauli blocking effects as well as the higher-order nucleon-nucleon (NN) 
correlations at large NM densities. Such medium effects serve as the physics 
origin of the empirical density dependence introduced into various versions 
of the effective NN interactions being widely used in the nuclear structure 
and reaction studies. Quite popular are the density dependent versions of the M3Y 
interaction (originally constructed in terms of three Yukawas to reproduce 
the G-matrix elements of the Reid \cite{Be77} and Paris \cite{An83} NN potentials 
in an oscillator basis), which have been applied successfully in the folding model
studies of nuclear scattering \cite{Kho96,Kho97,Kho07,Kho07r} or in the HF 
structure calculations of finite nuclei \cite{Na03,Na08,Na13}.  The different 
versions of the density dependent M3Y interaction have been used recently, together 
with the D1S and D1N versions of Gogny interaction \cite{Be91,Ch08} and 
SLy4 version \cite{Ch98} of Skyrme interaction \cite{Ch97}, to study the basic
properties of asymmetric NM at zero temperature \cite{Tha09} as well as 
the $\beta$-stable matter of cold neutron star \cite{Loa11}. While all 
these interactions give about the same description of the saturation 
properties of symmetric NM, the HF results for asymmetric NM \cite{Tha09} 
show that they are roughly divided into two groups that are associated 
with two different (so-called \emph{soft} and \emph{stiff}) behaviors 
of the nuclear symmetry energy at high NM densities. 
In the soft scenario, the symmetry energy changes its slope and decreases
at high NM densities, resulting on a drastic decrease of the
proton and lepton components in the core of neutron star that 
then becomes $\beta$-unstable. A very small proton fraction in neutron 
star matter given by the soft-type mean-field potentials excludes 
the direct Urca (DU) process in the neutron star cooling, whereas the DU process 
is well possible in the $\beta$-equilibrated neutron star matter predicted by the 
stiff-type mean field potentials \cite{Loa11}. A vital role of the nuclear symmetry 
energy has also been illustrated with the density dependent M3Y interaction, where 
a significant reduction of the maximum gravitational mass and radius of neutron 
star was found when the slope of the symmetry energy was changed from the stiff 
behavior to the soft one \cite{Loa11}.   

Because the composition and physical conditions of a hot newly born PNS 
are quite different from those of a cold and lepton-poor neutron star, it is 
of high interest to extend the mean-field approach of Ref.~\cite{Loa11} to the 
study of the hot baryon matter of PNS, and explore the sensitivity 
of the PNS composition to the nuclear symmetry energy at finite temperature and 
entropy as well as the effects caused by neutrino trapping. For this purpose,
the reliability of the present HF approach was tested first in the mean-field
study of hot and dense asymmetric NM, and the results are presented in 
Sec.~\ref{sec1}. These results turn out to be quite complementary to those
of the recent mean-field studies of thermal properties of asymmetric NM using 
different choices of the in-medium NN interaction \cite{Con14,Wel15,Con15} 
as well as the RMF studies \cite{Fed15,Hem15}. The HF results obtained for the EOS 
of hot $\beta$-stable PNS matter in both the $\nu$-free and $\nu$-trapped 
scenarios are presented in Sec.~\ref{sec2}, where the impact of the nuclear symmetry 
energy and nucleon effective mass on the hydrostatic configuration of hot PNS at the 
total entropy per baryon $S/A= 1,2$ and 4 was found very significant. The $\nu$-free 
PNS configuration at $S/A=4$ was studied in details in view of the results of  
hydrodynamic simulation by Hempel {\it et al.} \cite{Hem12}, which show that the 
neutrino-poor PNS matter with $S/A\approx 4$ occurs at the onset of a 40 
$M_\odot$ protonneutron progenitor collapse to black hole. The summary and main 
conclusions of the present research are given in Sec.~\ref{sec3}.            

\section{Hartree-Fock calculation of hot nuclear matter}
\label{sec1}
In our consistent HF approach, we consider a homogeneous spin-saturated NM over 
a wide range of the neutron and proton number densities, $n_n$ and $n_p$, or 
equivalently of the total nucleon number density $n_b=n_n+n_p$ (hereafter referred 
to as baryon density) and the neutron-proton asymmetry $\delta=(n_n-n_p)/(n_n+n_p)$. 
Given the direct ($v_{\rm D}$) and exchange ($v_{\rm EX}$) parts of the (central) 
in-medium NN interaction $v_{\rm c}$, the total energy density of NM at the given
baryon density and temperature $T$ is determined as
\begin{equation}
 E(T,n_b,\delta)=E_{\rm kin}(T,n_b,\delta)+E_{\rm pot}(T,n_b,\delta), \label{ek1}
\end{equation}
where the kinetic and HF potential energy densities are determined as
\begin{eqnarray}
 E_{\rm kin}(T,n_b,\delta)&=&\sum_{k \sigma \tau} n_{\sigma\tau}(\bm{k},T)
 \frac{\hbar^{2}k^2}{2m_\tau}  \label{ek2} \\
 E_{\rm pot}(T,n_b,\delta)&=&{\frac{1}{ 2}}\sum_{k \sigma \tau} \sum_{k'\sigma '\tau '}
 n_{\sigma\tau}(\bm{k},T)n_{\sigma'\tau'}(\bm{k}',T)
[\langle{\bm{k}\sigma\tau,\bm{k}'\sigma'\tau'}|v_{\rm D}|
{\bm{k}\sigma\tau,\bm{k}'\sigma'\tau'}\rangle \nonumber \\
& & \hskip 2cm +\ \langle{\bm{k}\sigma\tau,\bm{k}'\sigma'\tau'}|v_{\rm EX}|
{\bm{k}'\sigma\tau,\bm{k}\sigma'\tau'}\rangle] \label{ek2a} \\
&\equiv& {\frac{1}{ 2}}\sum_{k \sigma \tau} \sum_{k'\sigma '\tau '}
 n_{\sigma\tau}(\bm{k},T)n_{\sigma'\tau'}(\bm{k}',T)
\langle{\bm{k}\sigma\tau,\bm{k}'\sigma'\tau'}|v_{\rm c}|
{\bm{k}\sigma\tau,\bm{k}'\sigma'\tau'}\rangle_\mathcal{A}. \label{ek2b} 
\end{eqnarray}
The single-particle wave function $|\bm{k}\sigma\tau\rangle$ is plane wave, 
and the summation in Eqs.~(\ref{ek2})-(\ref{ek2b}) is done separately over 
the neutron ($\tau=n$) and proton ($\tau=p$) single-particle indices. The nucleon 
momentum distribution $n_{\sigma\tau}(\bm{k},T)$ in the hot, spin-saturated NM 
is given by the Fermi-Dirac distribution
\begin{equation}
  n_{\sigma\tau}(\bm{k},T)\equiv n_\tau(n_b,\bm k,\delta,T)=
 \frac{1}{1+\exp\{[\varepsilon_\tau(n_b,\bm k,\delta,T)-\mu_\tau]/T\}}. \label{ek3} 
\end{equation}
Here $T$ is the temperature (in MeV) and $\mu_\tau$ is the nucleon chemical 
potential. The single-particle energy $\varepsilon_\tau$ is given by
\begin{equation}
\varepsilon_\tau(n_b,\bm k,\delta,T)=\frac{\partial E(T)}
{\partial n_\tau(n_b,\bm k,\delta,T)}=\frac{\hbar^2 k^2}{2m_\tau}
+U_\tau(n_b,\bm k,\delta,T), \label{ek4}
\end{equation}
which is the change of the total NM energy caused by the removal or addition of a 
nucleon with the momentum $k$. At zero temperature, the nucleon momentum 
distribution (\ref{ek3}) is reduced to the step function determined with the 
Fermi momentum $k^{(\tau)}_F=(3\pi^2n_\tau)^{1/3}$  
\begin{eqnarray}
 n_{\tau}(n_b,k,\delta)=\left\{\begin{array}{ccc}
 1 &\mbox{if $k \leqslant k^{(\tau)}_F$} \\
 0 &\mbox{otherwise.} 
\end{array} \right. 
\end{eqnarray}
The single-particle potential $U_\tau$ consists of the HF term and 
rearrangement term (RT) 
\begin{eqnarray}
U_\tau(n_b,\bm k,\delta,T)&=& U^{\rm (HF)}_\tau(n_b,\bm k,\delta,T)+
U^{\rm (RT)}(n_b,\bm k,\delta,T), \label{uk} \\
{\rm where}\ \ U^{\rm (HF)}_\tau(n_b,\bm k,\delta,T) &=&\sum_{k'\sigma' \tau'}
n_{\tau'}(n_b,\bm{k}',\delta,T)\langle \bm{k}\sigma\tau,\bm{k}'\sigma'\tau'|v_{\rm c}|
\bm{k}\sigma\tau,\bm{k}'\sigma'\tau'\rangle_\mathcal{A} \label{uk1}\\
{\rm and}\ \ U^{\rm (RT)}_\tau(n_b,\bm k,\delta,T)&=&\frac{1}{2}
 \sum_{k_1\sigma_1\tau_1}\sum_{k_2\sigma_2\tau_2}n_{\tau_1}(n_b,\bm{k}_1,\delta,T)
 n_{\tau_2}(n_b,\bm{k}_2,\delta,T) \nonumber \\
 & & \ \times\left\langle \bm{k}_1\sigma_1\tau_1,\bm{k}_2\sigma_2\tau_2\left|
 \frac{\partial v_{\rm c}}{\partial n_\tau(n_b,\bm{k},\delta,T)}\right|
\bm{k}_1\sigma_1\tau_1,\bm{k}_2\sigma_2\tau_2\right\rangle_\mathcal{A}. \label{uk2}
\end{eqnarray}
At zero temperature, when the nucleon momentum approaches the Fermi momentum 
$(k\to k^{(\tau)}_F),\ \varepsilon_\tau(n_b,k^{(\tau)}_F,\delta,T=0)$ determined 
from Eqs.~(\ref{ek4})-(\ref{uk2}) is exactly the Fermi energy given by the 
Hugenholtz - van Hove theorem \cite{HvH}, which gives rise naturally to the 
RT of the single-particle potential if the in-medium NN interaction $v_{\rm c}$ 
is density dependent. At finite temperature $T$, the single-particle potential 
(\ref{uk}) is determined by an iterative procedure, with the chemical potential 
$\mu_\tau$ being determined at each iteration by normalizing the nucleon momentum 
distribution (\ref{ek3}) to the nucleon number density $n_\tau$   
\begin{equation}
 n_\tau = \frac{g}{(2\pi)^3}\int n_\tau(n_b,{\bm k},\delta,T) d{\bm k}, 
 \label{ek5}
\end{equation}
where $g=2$ is the spin degeneracy factor. The thermodynamic equilibrium 
of hot NM is directly associated with the evolution of the entropy. 
Given the NM energy density obtained microscopically (in terms of the nucleon 
degrees of freedom) in the HF calculation (\ref{ek1})-(\ref{ek2b}) using the 
realistic effective NN interaction between nucleons in medium, it is natural to 
determine the baryon entropy density $S$ of asymmetric NM on the same level, 
using the functional form for the entropy density of a Fermi-Dirac gas at the given 
temperature $T$ and baryon density $n_b$
\begin{eqnarray}
 S(T,n_b,\delta)&=&\frac{g}{8\pi^3}\sum_{\tau}\int\{n_\tau(n_b,{\bm k},\delta,T)
 \ln[n_\tau(n_b,{\bm k},\delta,T)] \nonumber \\
 && +\ [1-n_\tau(n_b,{\bm k},\delta,T)]\ln[1-n_\tau(n_b,{\bm k},\delta,T)]\}d{\bm k}. 
 \label{ek6}
\end{eqnarray}
Thus, the entropy density (\ref{ek6}) is affected by different nuclear EOS's 
via the corresponding nucleon mean-field potentials entering the nucleon 
momentum distribution (\ref{ek3}). 

The Helmholtz free energy density is determined as $F(T)=E(T)-TS(T)$. 
Dividing (\ref{ek1}) and (\ref{ek6}) over the total baryon number 
density $n_b$, we obtain the internal energy per particle $E(T)/A$ 
and the entropy per particle $S(T)/A$ that are used to determine the 
(density dependent) free energy per particle and the pressure of hot 
asymmetric NM as
\begin{equation}
\frac{F(T,n_b,\delta)}{A}=\frac{E(T,n_b,\delta)}{A}-
 T\frac{S(T,n_b,\delta)}{A},\ \ 
 P(T,n_b,\delta)=n_b^2\frac{\partial [F(T,n_b,\delta)/A]}
{\partial n_b}.  \label{ek7}  
\end{equation}
Thus, the EOS at finite temperature is given by Eq.~(\ref{ek7}). The 
different EOS's of symmetric NM are usually distinguished by the 
value of nuclear incompressibility $K_0$, determined at the nucleon
saturation density ($n_0\approx 0.17$\ fm$^{-3}$) and zero temperature as  
\begin{equation} 
 K_0=9n_b^2{\frac{\partial^2}
 {{\partial n_b^2}}}\left[\frac{E(T=0,n_b,\delta=0)}
 {A}\right]_{\displaystyle{n_b=n_0}}. \label {ek7k}
\end{equation}

Like in Ref.~\cite{Loa11}, we have used in the present HF calculation two 
different sets of the density-dependent M3Y interaction. The first set, dubbed 
as CDM3Yn (n=3,6), is the original M3Y-Paris interaction \cite{An83} 
supplemented by an \emph{isoscalar} (IS) density dependence that was parametrized 
\cite{Kho97} to reproduce the saturation properties of cold symmetric NM and give 
the different nuclear incompressibilities $K_0$ shown in Table~\ref{t1}. 
\begin{table}
\setlength{\tabcolsep}{0.2cm}
\renewcommand{\arraystretch}{0.7}
\caption{HF results for the NM saturation properties obtained at $n_b=n_0$
and $T=0$, using the considered effective NN interactions. The nucleon effective 
mass $m^*/m$ is evaluated at $\delta=0$ and $E_0=E(T=0,n_0,\delta=0)/A$. $K_{\rm sym}$ 
is the curvature parameter of the symmetry energy (\ref{ek10}), and $K_\tau$ is the 
symmetry term of the nuclear incompressibility (\ref{ek10k}) determined at the 
saturation density $n_\delta$ of asymmetric NM.}\vspace{0.5cm}
\centering\label{t1}
\begin{tabular}{cccccccccc} \hline
 Model & $n_0 $& $E_0$ & $K_0$ & $m^*/m$ & $J$ & $L$ & $K_{\rm sym} $&$K_\tau$ & Ref.\\ 
  & (fm$^{-3}$) & (MeV) & (MeV) & & (MeV) & (MeV) & (MeV) & (MeV) & \\ \hline
CDM3Y3  & 0.17 & -15.9 & 218 & 0.706 & 30.1 & 49.6 & -23& -181 & \cite{Loa15} \\ 
CDM3Y6  & 0.17 & -15.9 & 252 & 0.706 & 30.1 & 49.7 & -29 &-245 & \cite{Loa15} \\ 
CDM3Y3s  & 0.17 & -15.9 & 218 & 0.706 & 32.0 & 49.1 & -140 & -297 & \cite{Loa11} \\ 
CDM3Y6s  & 0.17 & -15.9 & 252 & 0.706 & 32.0 & 49.1 & -154 &-370 & \cite{Loa11} \\
M3Y-P5  & 0.16 & -16.1 & 235 & 0.637 & 30.9 & 27.9 & -229 &-339 & \cite{Na08}\\ 
M3Y-P7  & 0.16 & -16.0 & 254 & 0.589 & 33.0 & 54.3 & -138 &-395 & \cite{Na13} \\ 
D1N     & 0.16 & -16.0 & 221 & 0.748 & 30.1 & 32.4 & -182&-319 & \cite{Ch08}\\ 
SLy4    & 0.16 & -16.0 & 230 & 0.694 & 32.1 & 46.0 &-120 &-332 & \cite{Ch98} \\ \hline 
\end{tabular} 
\end{table}
These interactions, especially the CDM3Y6 version, have been widely tested in 
numerous folding model analyses of the nucleus-nucleus elastic scattering 
\cite{Kho07r} and charge-exchange scattering to the isobar analog states 
\cite{Kho07,Kho14}. The \emph{isovector} (IV) density dependence of the CDM3Yn 
interaction was first parametrized to reproduce the BHF results for the density- 
and isospin dependent nucleon optical potential of Jeukenne, Lejeune and Mahaux 
(JLM) \cite{Je77,Lej80}. Then, the total IV strength of the CDM3Yn interaction was 
fine-tuned to reproduce the measured charge-exchange $(p,n)$ and $(^3$He,$t)$ 
data for the isobar analog states \cite{Kho07,Kho14}. In our recent extended HF 
study of the single-particle potential in the NM \cite{Loa15}, the parameters of 
the IV density dependence of the CDM3Yn interactions have been redetermined 
with a consistent inclusion of the rearrangement term into the single-nucleon 
potential at different NM densities.   
At variance with the CDM3Yn interactions, the M3Y-Pn interactions (n=5 and 7)
have been carefully parametrized by Nakada \cite{Na03,Na08,Na13} in terms of the 
finite-range M3Y interaction supplemented with a zero-range density-dependent 
term, to consistently reproduce the NM saturation properties and ground-state 
(g.s.) structure of the stable nuclei as well as the dripline nuclei.
In particular, the newly parametrized M3Y-P7 version \cite{Na13} gives a stable 
description of symmetric NM up to the baryon density as high as 6$n_0$, while 
retaining a realistic description of the g.s. structure of finite nuclei.
For comparison, the present HF calculation has also been done using the two 
other popular choices of the effective NN interaction: the D1N version of Gogny 
interaction \cite{Ch08} and SLy4 version \cite{Ch97} of Skyrme interaction \cite{Ch98}.
The explicit parametrizations of the density dependent NN interactions used in 
the present HF calculation are presented and discussed in more details in 
Appendices A and B. The NM saturation properties at zero temperature given by 
different density dependent interactions are presented in Table~\ref{t1}. One can 
see that all interactions under study give reasonable description of the bulk 
properties of NM at the saturation density $n_0$, excepting the nucleon effective 
mass given by the M3Y-Pn interactions that is significantly lower than the 
empirical value around 0.72 \cite{Hs83}. 
\begin{figure}
\includegraphics[width=\textwidth]{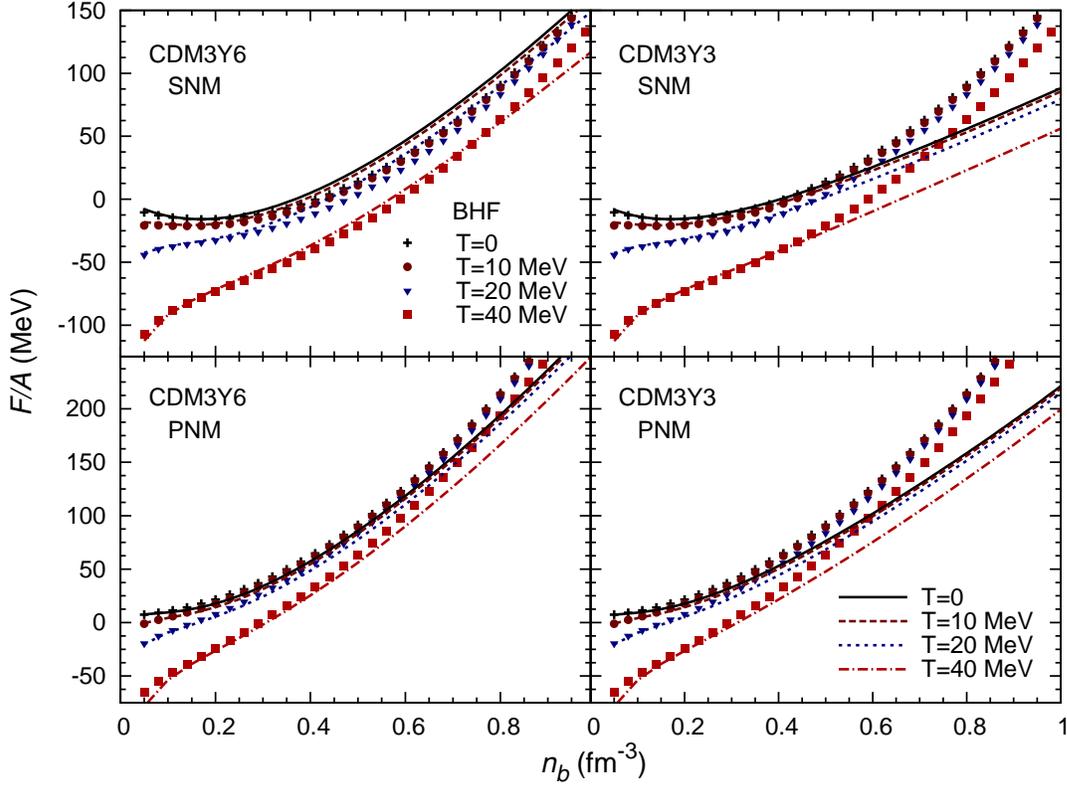}\vspace*{-1cm}
 \caption{(Color online) Free energy per particle $F/A$ of symmetric nuclear
matter (SNM) and pure neutron matter (PNM) at different temperatures given 
by the HF calculation (\ref{ek1})-(\ref{ek7}) using the CDM3Y3 (right panel) and 
CDM3Y6 (left panel) interactions \cite{Loa15} (lines), in comparison with the 
BHF results (symbols) by Burgio and Schulze \cite{Bur10}.} \label{f1}
\end{figure}
\begin{figure}
\includegraphics[width=\textwidth]{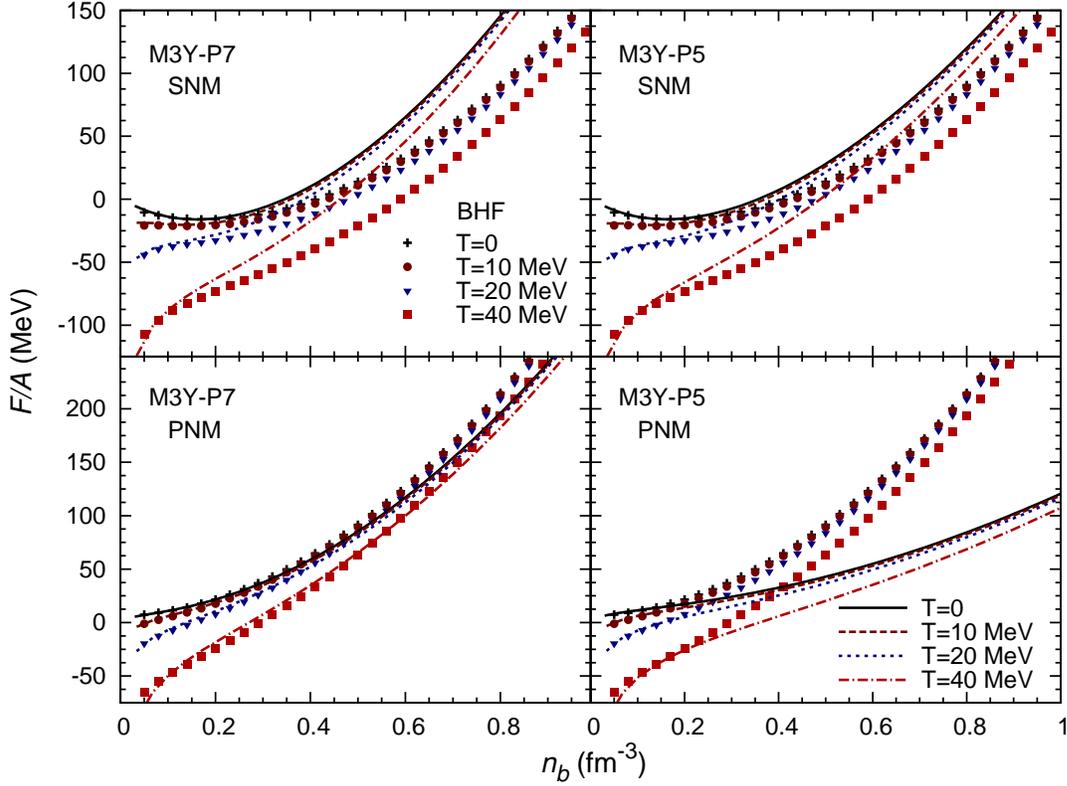}\vspace*{-1cm}
 \caption{(Color online) The same as Fig.~\ref{f1} but for the HF results 
 obtained with the M3Y-P5 (right panel) and M3Y-P7 (left panel) interactions 
 parametrized by Nakada \cite{Na08,Na13}.} \label{f2}
\end{figure}
\begin{figure}
\includegraphics[width=\textwidth]{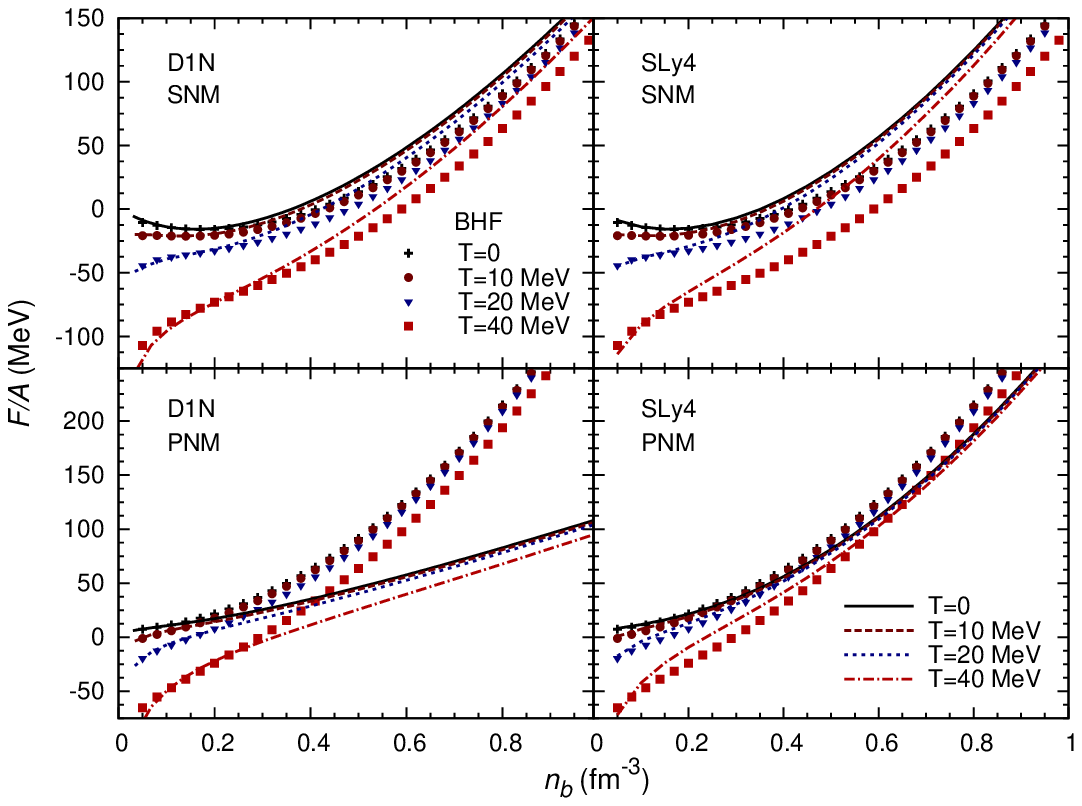}\vspace*{-1cm}
 \caption{(Color online) The same as Fig.~\ref{f1} but for the HF results 
obtained with the D1N version \cite{Ch08} of Gogny interaction (left panel)
and SLy4 version \cite{Ch98} of Skyrme interaction (right panel).} \label{f3}
\end{figure}
Given the HF mean field potentials based on various density dependent NN 
interactions (see Table~\ref{t1}), it is important to make some comparison of the 
present HF results with those of a microscopic many-body calculation starting from 
the realistic free NN interaction. In the present work we have focused on the 
microscopic EOS of hot asymmetric NM from the BHF calculation by Burgio and 
Schulze \cite{Bur10}, using the Argonne NN interaction (V18 version \cite{Wir95} 
with the Urbana three-body force). This microscopic EOS has been used to study the 
composition of hot $\beta$-stable PNS matter and estimate the maximum gravitational 
mass of PNS at entropy $S/A$=1 and 2. One can see in Fig.~\ref{f1} that the HF 
results obtained with the CDM3Y6 and CDM3Y3 interactions for the free energy 
per particle $F/A$ at different temperatures are in good agreement with the 
microscopic BHF results (reproduced here from the analytical expressions fitted 
by the authors of Ref.~\cite{Bur10}). A slight difference between the results 
given by the CDM3Y6 and CDM3Y3 interactions at high NM densities is mainly 
due to the different incompressibilities $K_0$ given by these two interactions. 
Over the whole range of NM densities up to $n_b=1$ fm$^{-3}$, the $F/A$ 
values given by the CDM3Y6 interaction are very close to the BHF results 
\cite{Bur10}. The $F/A$ values given by the M3Y-P5 and M3Y-P7 interactions 
are compared with the BHF results in Fig.~\ref{f2}, and a better agreement 
with the BHF results for pure neutron matter is achieved with the newly 
parametrized and improved M3Y-P7 version \cite{Na13}. A similar comparison of the 
free energy given by the D1N version \cite{Ch08} of Gogny interaction and SLy4 
version \cite{Ch98} of Skyrme interaction with the BHF results is shown in 
Fig.~\ref{f3}. We found that the results given by the D1N interaction for pure 
neutron matter differ strongly from the BHF results at high NM densities, while 
those given by the Sly4 interaction agree reasonably with the BHF results for 
both the symmetric NM and pure neutron matter over a wide range of NM densities.
Note that the free energy given by Sly4 interaction at high baryon densities
depends weakly on temperature because the internal NM energy obtained with the 
zero-range Skyrme interaction [see Appendix B] is \emph{temperature independent}.   

\subsection{The nuclear symmetry energy at finite temperature}\label{sec1.1} 
The difference between the free energy per particle of asymmetric NM and that 
of symmetric NM determines the \emph{free symmetry energy} per particle 
\begin{equation}
\frac{F_{\rm sym}(T,n_b,\delta)}{A}=\frac{F(T,n_b,\delta)}{A}
 -\frac{F(T,n_b,\delta=0)}{A}. \label{ek8}  
\end{equation}
\begin{figure}
\includegraphics[width=\textwidth]{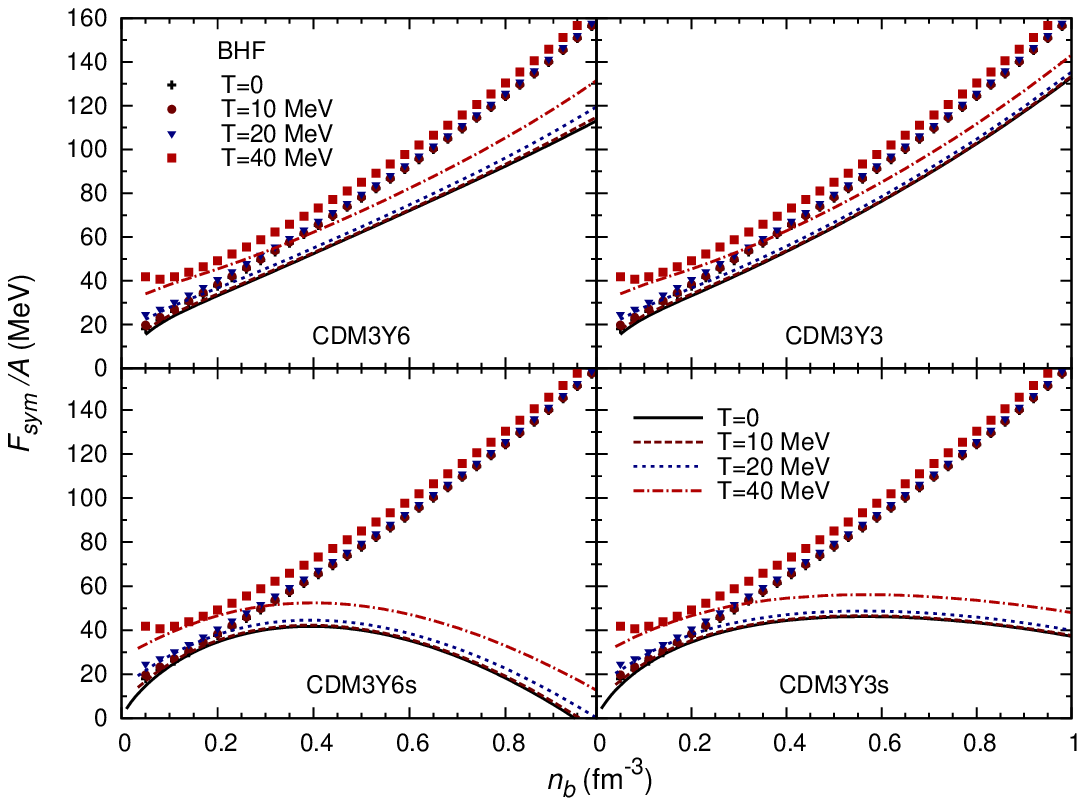}\vspace*{-1cm}
 \caption{(Color online) Free symmetry energy per particle $F_{\rm sym}/A$ 
of pure neutron matter ($\delta=1$) at different temperatures given 
by the HF calculations (\ref{ek8}) using the CDM3Y3 and CDM3Y6 interactions 
\cite{Loa15} and their soft versions CDM3Y3s and CDM3Y6s \cite{Loa11} 
(lines), in comparison with the BHF results (symbols) by Burgio and Schulze 
\cite{Bur10}.} \label{f4}
\end{figure}
\begin{figure}
\includegraphics[width=\textwidth]{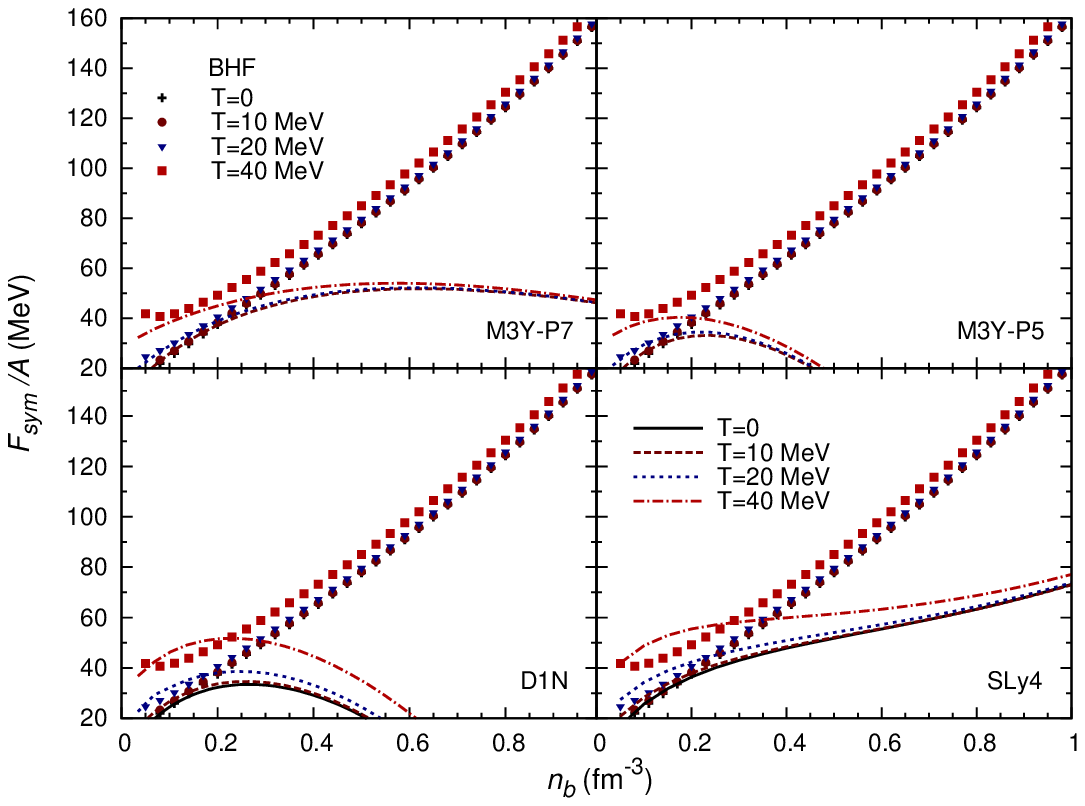}\vspace*{-1cm}
 \caption{(Color online) The same as Fig.~\ref{f4} but for the HF results 
given by the M3Y-P5 and M3Y-P7 interactions parametrized by Nakada \cite{Na08,Na13}, 
and the D1N version \cite{Ch08} of Gogny interaction and SLy4 version \cite{Ch98} 
of Skyrme interaction.} \label{f5}
\end{figure}    
It is often assumed \cite{Bur10,Li10}, like at the zero temperature, that 
the free symmetry energy per particle also depends \emph{quadratically} on the 
neutron-proton asymmetry $\delta$ as 
\begin{equation}
\frac{F_{\rm sym}(T,n_b,\delta)}{A}\approx 
 f_{\rm sym}(T,n_b)\delta^2+O(\delta^4), \label{ek9}
\end{equation}
and one needs only to determine the free energy of symmetric NM and that 
of pure neutron matter for the study of hot asymmetric NM. The free energy 
of asymmetric NM at any neutron-proton asymmetry $\delta$ can then be estimated 
using Eq.~(\ref{ek9}). However, as shown below, the quadratic approximation 
(\ref{ek9}) becomes much poorer with the increasing $\delta$ and $T$. In fact, 
the quadratic approximation is reasonable only for the internal symmetry energy  
\begin{equation}
\frac{E_{\rm sym}(T,n_b,\delta)}{A}=\frac{E(T,n_b,\delta)}{A}
 -\frac{E(T,n_b,\delta=0)}{A}\approx 
\varepsilon_{\rm sym}(T,n_b)\delta^2+O(\delta^4).  \label{ek9p}  
\end{equation}
In the thermodynamic equilibrium, it is illustrative to express the free 
symmetry energy per particle as
\begin{equation}
\frac{F_{\rm sym}(T,n_b,\delta)}{A}=\frac{E_{\rm sym}(T,n_b,\delta)}{A}
 -T\frac{S_{\rm sym}(T,n_b,\delta)}{A}, \label{ek9k}  
\end{equation}
where $S_{\rm sym}(T,n_b,\delta)/A$ is the symmetry part of the entropy
per particle 
\begin{equation}
\frac{S_{\rm sym}(T,n_b,\delta)}{A}=\frac{S(T,n_b,\delta)}{A}
 -\frac{S(T,n_b,\delta=0)}{A} \label{ek9s}
\end{equation}      
A comparison of the HF results for the free symmetry energy with those 
given by the microscopic BHF calculation (reproduced from the fitted analytical 
expression given in Ref.~\cite{Bur10}) is much more drastic, because $F_{\rm sym}/A$ 
is directly determined by the isospin dependence of the considered NN interactions. 
As can be seen in Fig.~\ref{f4}, the free symmetry energy predicted by the HF
calculation using the CDM3Y6 and CDM3Y3 interactions agree well with the BHF 
results. Because the isovector density dependence of these two interactions 
has been fine tuned \cite{Kho07,Loa15} to reproduce the density dependence 
of the nucleon optical potential obtained in the BHF calculation by the JLM group 
\cite{Je77}, both the CDM3Y3 and CDM3Y6 interactions give a \emph{stiff} density
dependence of the free symmetry energy at different temperatures. To illustrate 
this we also used the \emph{soft} versions (CDM3Y3s and CDM3Y6s) of these interactions, 
where the IV density dependence $F_1(n_b)$ was assumed to be the same as the IS 
density dependence $F_0(n_b)$ \cite{Loa11}. One can see that the soft free symmetry 
energy fails to agree with the BHF results soon after the baryon density exceeds the 
saturation density $n_0$ (see lower panel of Fig.~\ref{f4}). 
\begin{figure}
\includegraphics[width=1.0\textwidth]{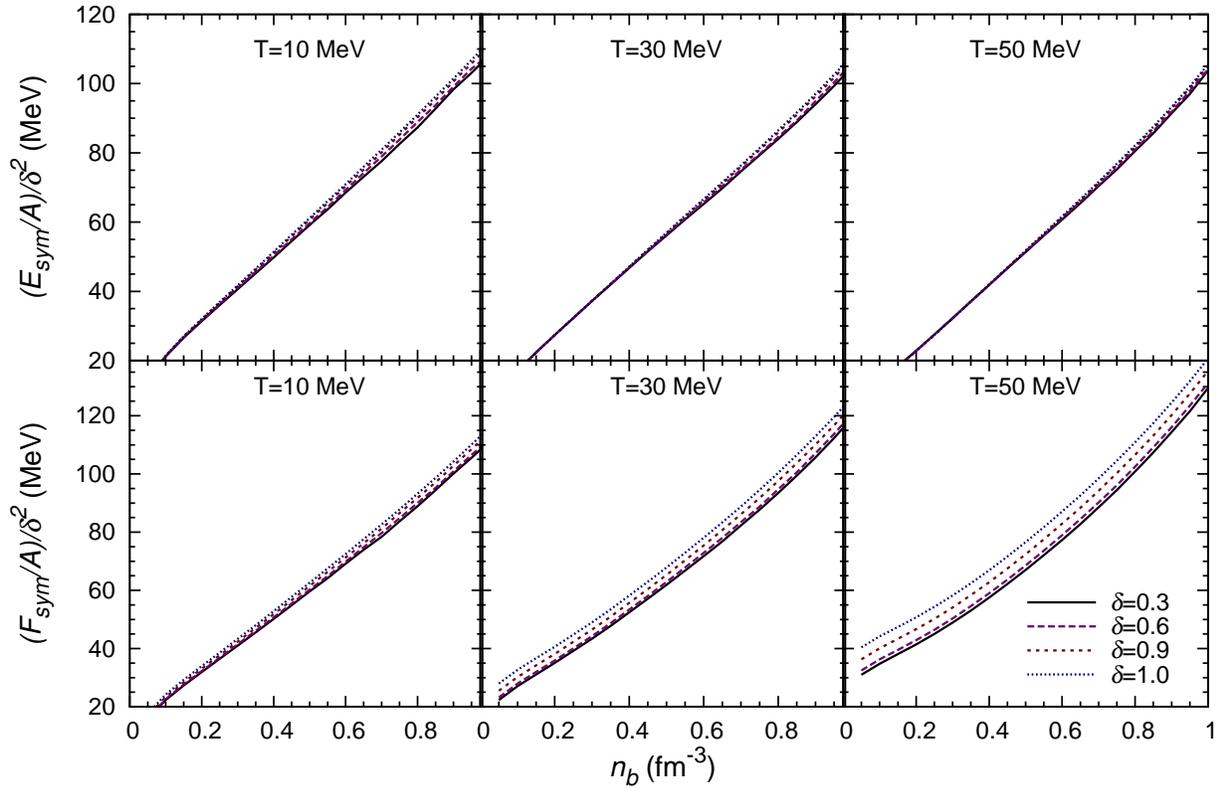}\vspace*{-1cm}
 \caption{(Color online) Free symmetry energy (\ref{ek8}) (lower panel) and  
internal symmetry energy (\ref{ek9p}) (upper panel) at different temperatures, 
given by the HF calculation using the CDM3Y6 interaction \cite{Loa15}.
$(F_{\rm sym}/A)/\delta^2$ curves must be very close if the quadratic 
approximation (\ref{ek9}) is valid.} \label{f6}
\end{figure}
The free symmetry energy given by the M3Y-Pn, D1N, and SLy4 interactions also 
disagree strongly with the BHF results (see Fig.~\ref{f5}) at high NM densities, 
although the isospin dependence of these interactions was carefully tailored 
to give the good description of the g.s. structure of the neutron-rich dripline nuclei 
\cite{Na08,Na13,Ch08}. These two different behaviors have been observed earlier 
\cite{Tha09} and are discussed in the literature as the \emph{asy-stiff} (with the 
symmetry energy steadily increasing with the NM density) and \emph{asy-soft} (with 
the symmetry energy reaching its saturation and then decreasing to negative values 
at high NM densities) behaviors. In this context, the CDM3Y3, CDM3Y6, and Sly4 
interactions are asy-stiff, while the CDM3Y3s, CDM3Y6s, M3Y-P5, M3Y-P7, and D1N 
interactions are asy-soft. It can be seen in Figs.~\ref{f4} and \ref{f5} that 
the finite-temperature effect does not change this classification. It should be
noted that, like the BHF calculation, the most recent nonrelativistic mean-field 
studies of hot asymmetric NM \cite{Con14,Wel15} as well as the RMF studies 
\cite{Fed15,Hem15} also predict the asy-stiff behavior of the free symmetry energy.

We have further checked the dependence of the free symmetry energy on the 
neutron-proton asymmetry $\delta$ by performing finite-temperature HF calculations 
at different $\delta$ values, and it turned out that the quadratic approximation 
(\ref{ek9}) becomes much poorer with the increasing temperature as shown in lower 
panel of Fig.~\ref{f6}, where all $(F_{\rm sym}/A)/\delta^2$ curves should be 
very close if the quadratic approximation (\ref{ek9}) is valid. Such a breakdown 
of the quadratic approximation, sometimes also dubbed as the parabolic law, is 
clearly due to the effect of finite entropy. Therefore, the free symmetry energy 
of hot asymmetric NM needs to be calculated explicitly at different neutron-proton 
asymmetries $\delta$ without using the approximation (\ref{ek9}). To further 
illustrate this effect, we have plotted the HF results for the internal symmetry 
energy per particle $E_{\rm sym}/A$ in the upper panel of Fig.~\ref{f6}, and one 
can see that the quadratic approximation (\ref{ek9p}) remains reasonable for 
$E_{\rm sym}/A$ at different temperatures.   

At zero temperature, $\varepsilon_{\rm sym}$ determined at the saturation 
density $n_b=n_0$ is known as the symmetry energy coefficient $J$ which has been 
predicted by numerous many-body calculations to be around 30 MeV. 
The symmetry energy of cold NM can be expanded around $n_0$ \cite{Tsa09,Pie09} as
\begin{equation}
 \varepsilon_{\rm sym}(T=0,n_b)=J+\frac{L}{3}\left(\frac{n_b-n_0}
 {n_0}\right) +\frac{K_{\rm sym}}{18}\left(\frac{n_b-n_0}{n_0}\right)^2
 + ... \label{ek10}
\end{equation}
where $L$ and $K_{\rm sym}$ are the slope and curvature parameters of the
symmetry energy. The asymmetric NM with the neutron-proton asymmetry 
$\delta\lesssim 0.75$ can still be saturated by the nuclear mean field, 
but at the baryon number density siginificantly lower than that of symmetric NM
($n_\delta<n_0$) \cite{Kho96}. As shown by Piekarewicz and Centelles \cite{Pie09},
the nuclear incompressibility at the new saturation density $n_\delta$ can be 
approximated by a quadratic dependence on the neutron-proton asymmetry $\delta$ 
\begin{equation}
 K(n_\delta)\approx K_0+K_\tau\delta^2+O(\delta^4). \label{ek10k}  
\end{equation}
We have calculated the symmetry term $K_\tau$ of the nuclear incompressibility 
$K(n_\delta)$ based on Eq.~(18c) in Ref.~\cite{Pie09}, using different density 
dependent NN interactions under the present study. The obtained $K_\tau$ values
of -180 to -395 MeV (see Table~\ref{t1}) turned out to be smaller (in $|K_\tau|$ 
value) than $K^{\rm (exp)}_\tau\approx -550\pm 100$ MeV deduced from the 
measured strength distribution of giant monopole resonance (GMR) in neutron-rich 
tin isotopes \cite{Li07}. Such a disagreement was found also with different versions 
of Skyrme interaction \cite{Li07} and FSUGold interaction that was well tested 
in the RMF studies of GMR in heavy nuclei \cite{Pie09}. This remains an open 
question that might be linked to large uncertainties in extrapolating $K_\tau$ 
values for asymmetric NM from the GMR data measured for finite  nuclei \cite{Pie09}.
 
The knowledge about the density dependence of the nuclear 
symmetry energy is vital for the construction of the EOS of asymmetric NM.
The widely used method to probe the NM symmetry energy associated (in the 
mean-field scheme) with a given in-medium NN interaction is to test 
such an interaction in the nuclear reaction and/or the nuclear structure studies 
involving nuclei with the large neutron excess. Based on the physics constraints 
implied by such studies, the extrapolation is made to draw the conclusion on the 
low- and high-density behavior of the symmetry energy  (see a recent review 
\cite{Ba14} for more details). For example, the parameters $J,L$, and 
$K_{\rm sym}$ of the expansion (\ref{ek10}) have been the subject of several
nuclear structure studies, where one tries to deduce constraints for these
3 parameters from the observation of the monopole, dipole, and spin-dipole
resonances \cite{Col14}. Note that higher-order term $O(\delta^4)$ in the 
expansion (\ref{ek9p}) of the symmetry energy at zero temperature might be 
needed for a more precise determination of the slope parameter $L$ \cite{Xu14}.
Quite a robust constraint for both the $J$ and $L$ values has been established 
\cite{Li13} based on several tens analyses of the terrestrial nuclear physics 
experiments and astrophysical observations, which give $J\approx 31.6\pm 2.7$ MeV 
and $L\approx 58.9\pm 16.0$ MeV. Using the parameters of the isovector density 
dependence of the CDM3Yn interactions determined recently in Ref.~\cite{Loa15}, 
the values $J\approx 30.1$ MeV and $L\approx 49.7$ MeV given by the HF 
calculation of cold asymmetric NM are well within this empirical range. In our 
previous HF study \cite{Loa11} of cold ($\beta$-stable) neutron star matter, 
the baryon and lepton compositions in the core of neutron star as well as its 
maximum gravitational mass and radius were shown to be strongly affected by the 
slope of the nuclear symmetry energy at high baryon densities. The extension
of the mean-field approach of Ref.~\cite{Loa11} to study the impact of the symmetry
energy to the composition of hot $\beta$-stable PNS matter is presented in the 
next Section.  

\subsection{The nucleon effective mass and thermal properties of NM}\label{sec1.2}
\begin{figure}
\includegraphics[width=0.9\textwidth]{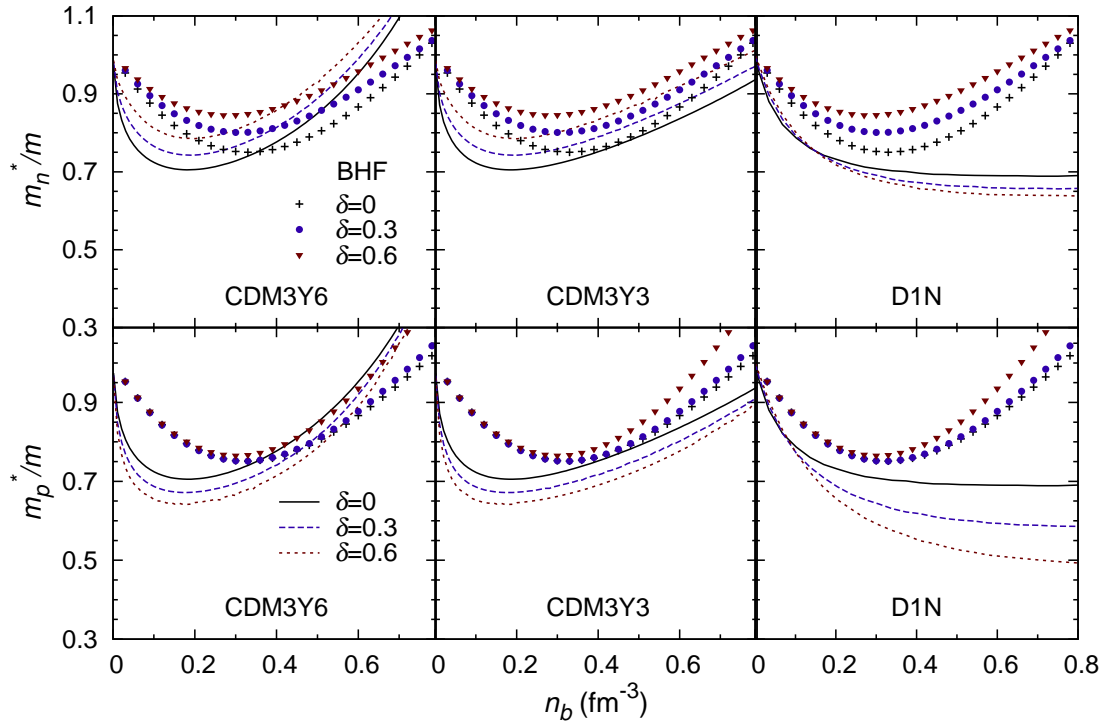}\vspace*{-0.5cm}
 \caption{(Color online) Density profile of the neutron- (upper panel) and  
proton effective mass (lower panel) at different neutron-proton asymmetries
$\delta$ given by the HF calculation using the CDM3Y6 and CDM3Y3 interactions 
\cite{Loa15}, and the D1N version of Gogny interaction \cite{Ch08}, 
in comparison with the BHF results (symbols) by Baldo {\it et al.} \cite{Bal14}.} 
\label{f1n}
\end{figure}
A very important physics quantity associated with the momentum dependence
of the single-nucleon potential (\ref{uk}) is the (density dependent)
nucleon effective mass, determined within the nonrelativistic mean-field formalism as
\begin{equation}
\frac{m^*_\tau(n_b,\delta)}{m}=\Biggl[1+\frac{m}{\hbar^2k^{(\tau)}_F}
\frac{\partial U_\tau(n_b,\delta,k)}{\partial k}\Bigg|_{k^{(\tau)}_F}\Biggr]^{-1}, 
 \label{eff1}
\end{equation}
where $m$ is the free nucleon mass. Given the finite range of the Yukawa (or Gaussian) 
function used to construct the radial part of the CDM3Yn, M3Y-Pn, and Gogny 
interactions, the single-particle potential depends explicitly on the nucleon 
momentum $k$ via its exchange term [see Eqs.~(\ref{apu3}) and (\ref{apu7}) 
in Appendix A] that implies a \emph{nonlocal} single-particle potential in the 
coordinate space. Therefore, the nucleon effective mass $m^*$ is a measure of  
nonlocality of the mean-field potential felt by a nucleon propagating in nuclear 
medium. Because of the zero range of Skyrme interaction, the corresponding 
mean-field potential is always local, and the momentum dependence of Skyrme 
interaction has to be included explicitly [see Eq.~(\ref{sk1}) in Appendix B]. 
In general, the nucleon effective mass is linked closely to several important nuclear 
physics phenomena, like the dynamics of HI collisions, damping of giant resonances, 
temperature profile and cooling of hot PNS and neutrino emission therefrom 
\cite{Pra97,Bal14,Li15}. Moreover, the density dependence of nucleon effective 
mass was shown to be directly related to the thermodynamic properties of NM 
\cite{Pra97,Ste00}. For example, in the degenerate limit of symmetric NM or pure 
neutron matter with $T\ll T_F$, the entropy per particle and temperature at the 
given Fermi momentum $k_F$ can be approximately expressed \cite{Pra97} as
\begin{equation}
 \frac{S}{A}\approx\frac{\pi^2}{2}\frac{T}{T_F},\ {\rm where\ the\ Fermi\ temperature}\ 
 T_F=\frac{\hbar^2k_F^2}{2m^*}. \label{eff2}
\end{equation}
\begin{figure}
\includegraphics[width=0.9\textwidth]{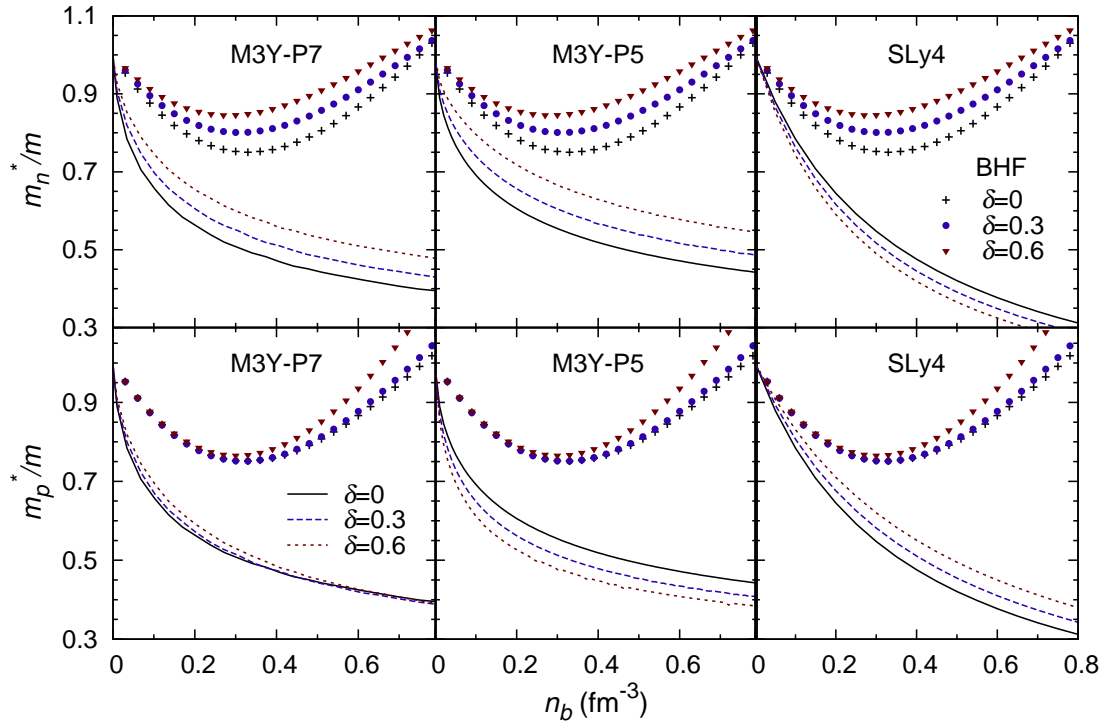}\vspace*{-0.5cm}
 \caption{(Color online) The same as Fig.~\ref{f1n} but for the HF results 
obtained with the M3Y-P7 and M3Y-P5 interactions \cite{Na08,Na13}, and 
the SLy4 version \cite{Ch98} of Skyrme interaction.} 
\label{f2n}
\end{figure}  

For a comparison with the BHF results for the nucleon effective mass at different 
NM densities \cite{Bal14}, we have evaluated the neutron- and proton effective 
mass (\ref{eff1}) from the HF single-particle potential at zero temperature. The 
density dependence of the neutron- and proton effective mass at different 
neutron-proton asymmetries $\delta$ given by different density dependent NN 
interactions are compared with the BHF results in Figs.~\ref{f1n} and \ref{f2n}. 
To be consistent with the discussion on results shown in Figs.~\ref{f1}-\ref{f5}, 
we have plotted in Figs.~\ref{f1n} and \ref{f2n} the results of the BHF calculation 
by Baldo {\it et al.} \cite{Bal14} (reproduced here from the analytical formulas 
fitted by the authors of Ref.~\cite{Bal14}) which were also obtained with the 
V18 version of Argonne NN interaction \cite{Wir95} supplemented by the 
phenomenological Urbana three-body force. We found that only the CDM3Yn interactions 
give the neutron- and proton effective mass behaving similarly to that predicted 
by the BHF calculation. Namely, the $m_\tau^*/m$ value decreases to some minimum 
at the baryon density $n_b\approx 0.2$ fm$^{-3}$ and then rises up to above unity 
at high baryon densities. Such a \emph{saturation} trend of the nucleon effective 
mass was shown in the BHF calculation to be entirely due to the repulsive 
contribution by the three-body force (see Fig.~1 of Ref.~\cite{Bal14}). The rise 
at high densities of the $m_\tau^*/m$ value given by the CDM3Y6 interaction is 
stiffer than that given by the CDM3Y3 interaction, and this effect (observed at
both $\delta=0$ and $\delta\neq 0$) is mainly due to different nuclear incompressibilities 
$K_0$ of cold symmetric NM determined with different IS density dependence $F_0(n_b)$ 
of these interactions [see Eq.~(\ref{app5}) in Appendix A]. The D1N 
version of Gogny interaction gives a rather slowly decreasing $m_\tau^*/m$ value 
with the increasing baryon density (see right panel of Fig.~\ref{f1n}), and such
a behavior of the nucleon effective mass is similar to that obtained in the recent 
mean-field studies using other choices of the finite-range effective NN interaction 
\cite{Con15}. The nucleon effective mass predicted by the M3Y-Pn and Sly4 
interactions (see Fig~\ref{f2n}) fall rapidly from $m_\tau^*/m\approx 0.6$ at 
$n_b\approx 0.2$ fm$^{-3}$ to low values around 0.3 at $n_b\approx 0.8$ fm$^{-3}$. 
Such a behavior seems not realistic and totally opposite to that predicted by the 
microscopic BHF calculation 
\cite{Bal14}. 
\begin{figure}
\includegraphics[width=0.9\textwidth]{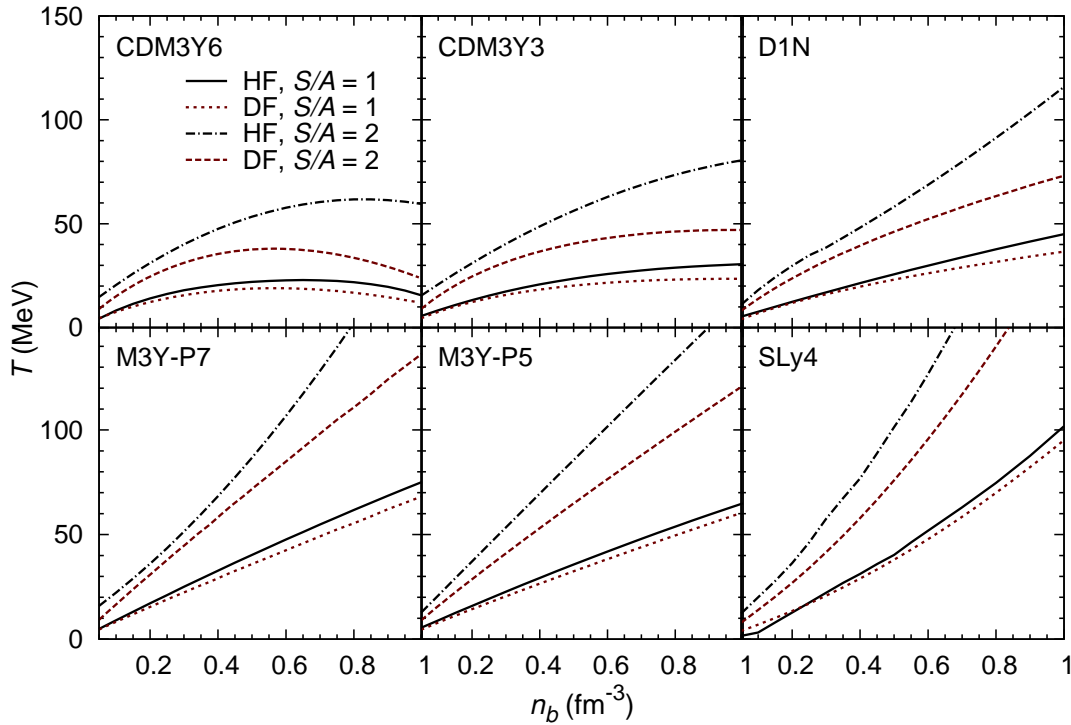}\vspace*{-0.5cm}
 \caption{(Color online) Density profile of temperature in the isentropic and 
symmetric NM given by the HF calculation using different density dependent NN 
interactions, in comparison with that given by the approximation (\ref{eff2}) for the 
fully degenerate Fermi (DF) system at $T\ll T_F$.} \label{f3n}
\end{figure}  
    
\begin{figure}
\includegraphics[width=1.0\textwidth]{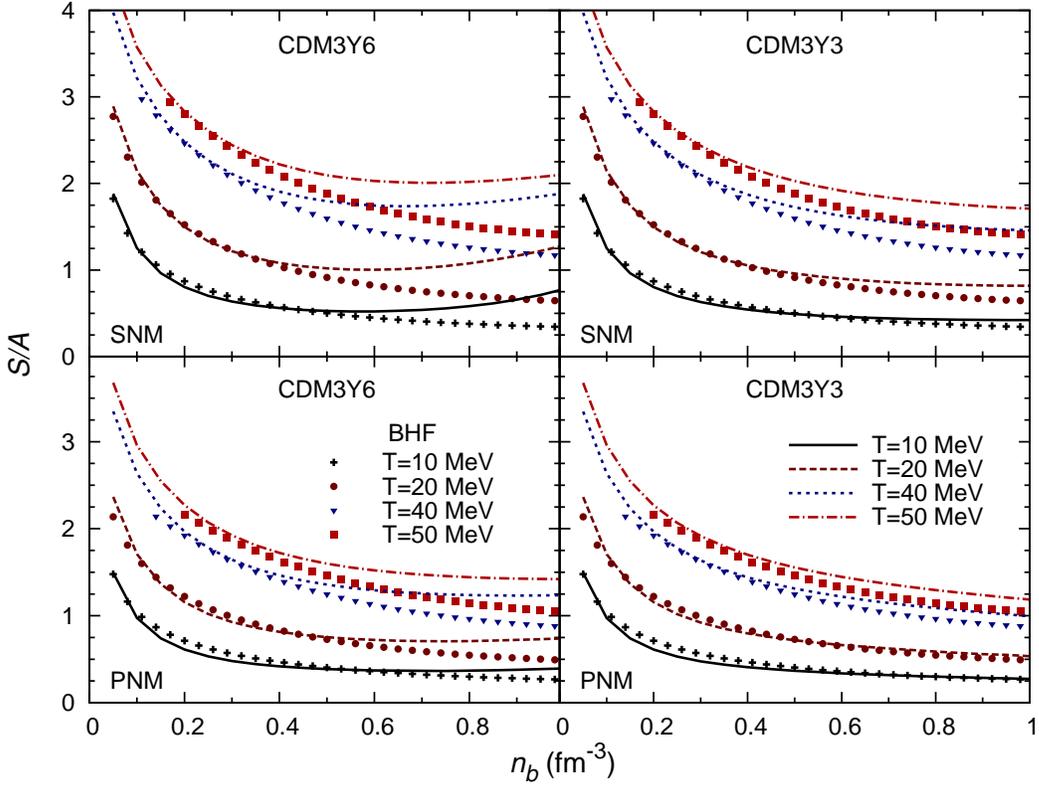}\vspace*{-0.5cm}
 \caption{(Color online) Density profile of entropy per particle $S/A$ of  
symmetric nuclear matter (SNM) and pure neutron matter (PNM) at different 
temperatures, deduced from the HF results (lines) obtained with the CDM3Y6 
and CDM3Y3 interactions \cite{Loa15}, in comparison with the BHF results by 
Burgio and Schulze (symbols) \cite{Bur10}.} \label{f7}
\end{figure}
\begin{figure}
\includegraphics[width=1.0\textwidth]{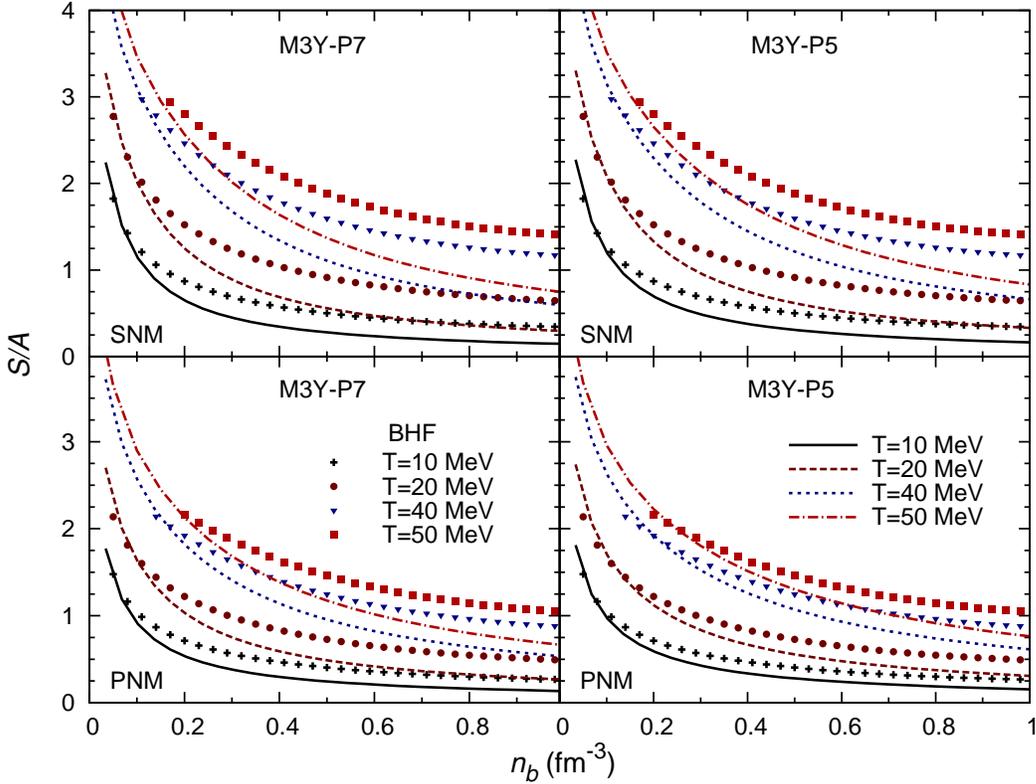}\vspace*{-0.5cm}
 \caption{(Color online) The same as Fig.~\ref{f7} but for the HF results 
obtained with the M3Y-P7 and M3Y-P5 interactions \cite{Na08,Na13}.} \label{f8}
\end{figure}
\begin{figure}
\includegraphics[width=1.0\textwidth]{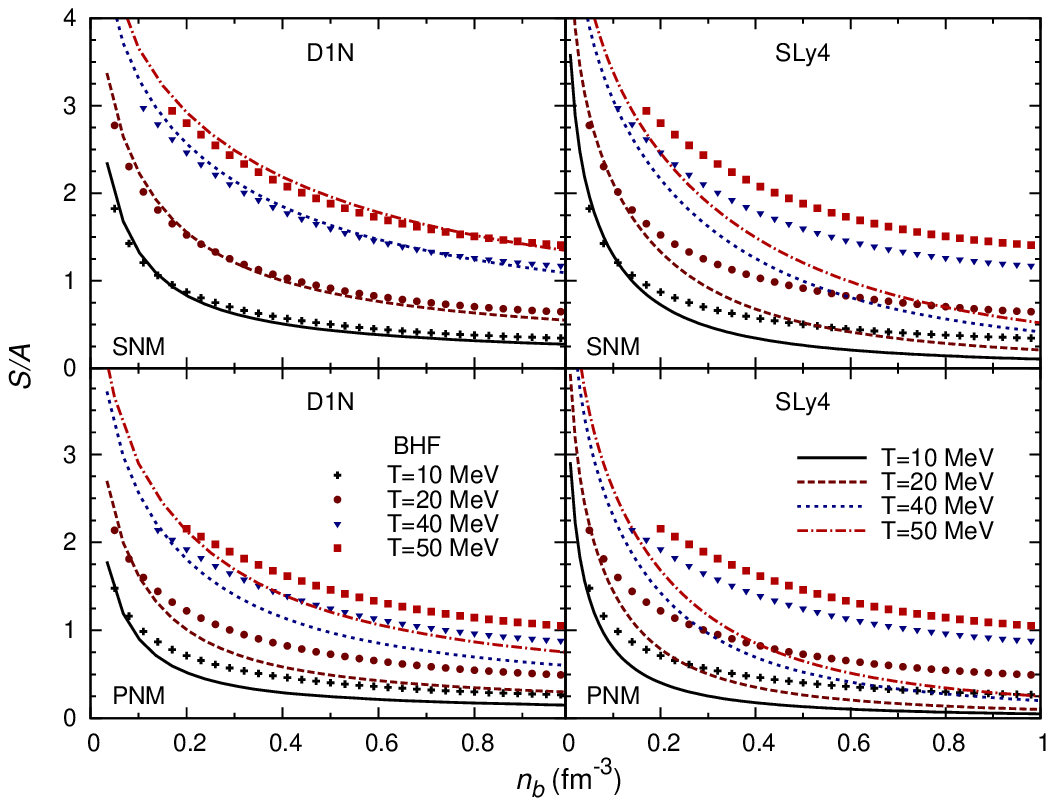}\vspace*{-0.5cm}
 \caption{(Color online) The same as Fig.~\ref{f7} but for the HF results 
obtained with the D1N version \cite{Ch08} of Gogny interaction and 
SLy4 version \cite{Ch98} of Skyrme interaction.} 
\label{f9}
\end{figure}
The difference in nucleon effective masses given by different density dependent
NN interactions shown in Figs.~\ref{f1n} and \ref{f2n} is very significant at high 
baryon densities, and it must show up, therefore, also in the HF results for
thermal properties of NM \cite{Pra97,Ste00}. As an illustration, the density 
profile of temperature of the isentropic and symmetric NM at $S/A=1$ and 2 
deduced from the HF results given by different density dependent NN interactions 
is compared with that given by the full degeneracy limit (\ref{eff2}) in 
Fig.~\ref{f3n}. One can see that at low temperatures (or $S/A=1$) the HF results 
are rather close to those given by Eq.~(\ref{eff2}). At $S/A=2$ the temperature 
predicted by the HF calculation is larger than that given by the degeneracy limit 
over the whole density range, and the approximation (\ref{eff2}) becomes worse 
at high baryon densities. Nevertheless, the trend of temperature $T$ being 
proportional to the Fermi temperature $T_F$ (or inversely proportional to nucleon 
effective mass) seems to hold, and at the given baryon density the lower the nucleon 
effective mass $m^*$ the higher the NM temperature $T$. Given very low nucleon 
effective masses at high baryon densities predicted by the M3Y-Pn and Sly4 
interactions compared to those predicted by the CDM3Yn and D1N interactions 
(see Fig~\ref{f2n}), the NM temperatures predicted by the M3Y-Pn and Sly4 interactions 
are also substantially higher than those predicted by the CDM3Yn and D1N interactions
at high baryon densities (see Fig~\ref{f3n}). This result shows clearly a strong 
impact of the nucleon effective mass to the thermodynamic properties of NM.    
 
The behavior of entropy is also an important thermodynamic property associated 
with the given EOS of hot NM (through the single-particle potential embedded in 
the nucleon momentum distribution used to estimate $S/A$). Although a constant 
$S/A=1\sim 2$ was often assumed for the hydrostatic configuration of PNS \cite{Pra97}, 
a recent hydrodynamic simulation of black hole formation in a failed core-collapse 
supernova \cite{Hem12} has shown that entropy per baryon $S/A\approx 4$ is reached 
at the onset of collapse of the massive ($40~M_\odot$) PNS to black hole. Therefore, 
the present HF calculation of hot NM has been extended to high temperatures $T$ that 
correspond to $S/A=4$. At $T\leqslant 50$ MeV, the density profiles of entropy 
given by the present HF calculation are compared in Figs.~\ref{f7}-\ref{f9} with 
those given by the BHF calculation \cite{Bur10} of symmetric NM and pure neutron 
matter (reproduced here from the fitted analytical expressions of the free 
energy in the temperature range $0\leqslant T\leqslant 50$ MeV \cite{Bur10}).  

\begin{figure}
\includegraphics[width=1.0\textwidth]{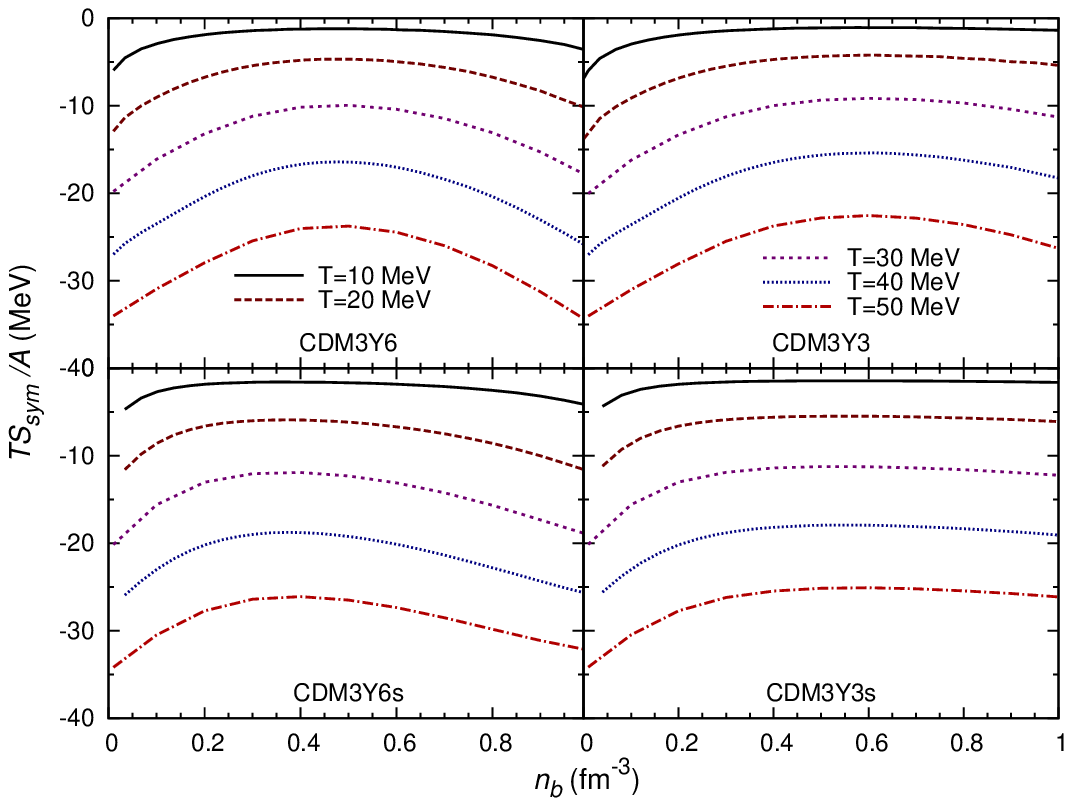}\vspace*{-0.5cm}
 \caption{(Color online) Symmetry part of the entropy per particle (\ref{ek9s}) 
of pure neutron matter at different temperatures given by the CDM3Y3 
and CDM3Y6 interactions \cite{Loa15} and their soft CDM3Y3s and CDM3Y6s 
versions \cite{Loa11}. $S_{\rm sym}/A$ is scaled by the corresponding temperature
to have the curves well distinguishable at different $T$.} \label{f10}
\end{figure}
One can see in Fig.~\ref{f7} that $S/A$ values obtained with both the CDM3Y3 
and CDM3Y6 interactions agree reasonably well with those given by the BHF calculation 
at different NM densities. At high densities up to $n_b=1$ fm$^{-3}$, the $S/A$ 
values given by the CDM3Y3 interaction agree better with those given by the BHF 
calculation \cite{Bur10} (see right panel of Fig.~\ref{f7}). Such a difference in 
the calculated $S/A$ values is directly resulted from the difference in the free
energy per particle $F/A$ (see Fig.~\ref{f1}), because entropy at the given baryon
density can also be determined from the derivative of the free energy with respect to 
the temperature, instead of using Eq.~(\ref{ek6}). Therefore, the different 
behaviors of $S/A$ values obtained with the two CDM3Yn interactions (found at 
both $\delta=0$ and $\delta=1$) are due mainly to different nuclear 
incompressibilities $K_0$ given by these two interactions, like discussed above
for the behavior of nucleon effective mass. Recall that the CDM3Y3 and CDM3Y6 
interactions give $K_0\approx 217$ and 252 MeV, respectively, compared with 
$K_0\approx 210$ MeV given by the BHF calculation \cite{Bur10}. The density 
profile of $S/A$ values given by the other interactions are plotted in 
Figs.~\ref{f8} and \ref{f9}, and the agreement with the BHF results 
becomes worse, especially for the Sly4 version of Skyrme interaction. The 
difference between $S/A$ values given by the HF calculation using Sly4 interaction 
and those given by the BHF calculation is likely due to the drastic difference 
in the nucleon effective masses at high baryon densities (see right panel of 
Fig.~\ref{f2n}) which implies different momentum dependences of the single-particle 
potential (the key input for the HF calculation of thermal properties of NM). 
To explore how strong is the effect caused by different behaviors of the nuclear 
symmetry energy, we have plotted in Fig.~\ref{f10} the symmetry part of entropy 
per particle (in terms of $TS_{\rm sym}/A$) of pure neutron matter at different 
temperatures, given by the CDM3Y3 and CDM3Y6 interactions \cite{Kho97,Loa15} and 
their soft CDM3Y3s and CDM3Y6s versions \cite{Loa11}. These two groups of the CDM3Yn 
interactions were shown in Fig.~\ref{f4} to give very different behaviors of the 
free symmetry energy at high baryon densities. As can be seen in Fig.~\ref{f10}, 
the maximal $TS_{\rm sym}/A$ values associated with the stiff symmetry energy 
is slightly higher than that associated with the soft symmetry energy. However, 
the stiff-soft difference shown in Fig.~\ref{f10} is not as drastic as that found 
in the free symmetry energy given by the CDM3Yn interactions and their soft versions 
at different temperatures (see Fig.~\ref{f4}). 

In conclusion, the different behaviors of entropy with the increasing baryon density 
shown in Figs.~\ref{f7}-\ref{f9} is due to difference in the momentum dependence
of single-particle potential used to construct the nucleon momentum distribution 
(\ref{ek3}), the main input for the determination of entropy (\ref{ek6}). 
This is also the source of difference in the nucleon effective mass obtained with 
different density dependent NN interactions shown in Figs.~\ref{f1n}-\ref{f2n}. 
In the same manner, a slight difference shown in Fig.~\ref{f10} for the symmetry part 
of entropy per particle should be due to different the momentum dependences 
of the IV term of single-particle potential obtained with the CDM3Yn interactions 
and their soft CDM3Yns versions. This is directly related to the difference 
between the stiff- and soft symmetry-energy scenarios.

\section{EOS of hot $\beta$-stable PNS matter}\label{sec2}
\begin{figure}
\includegraphics[angle=-90,width=1.0\textwidth] {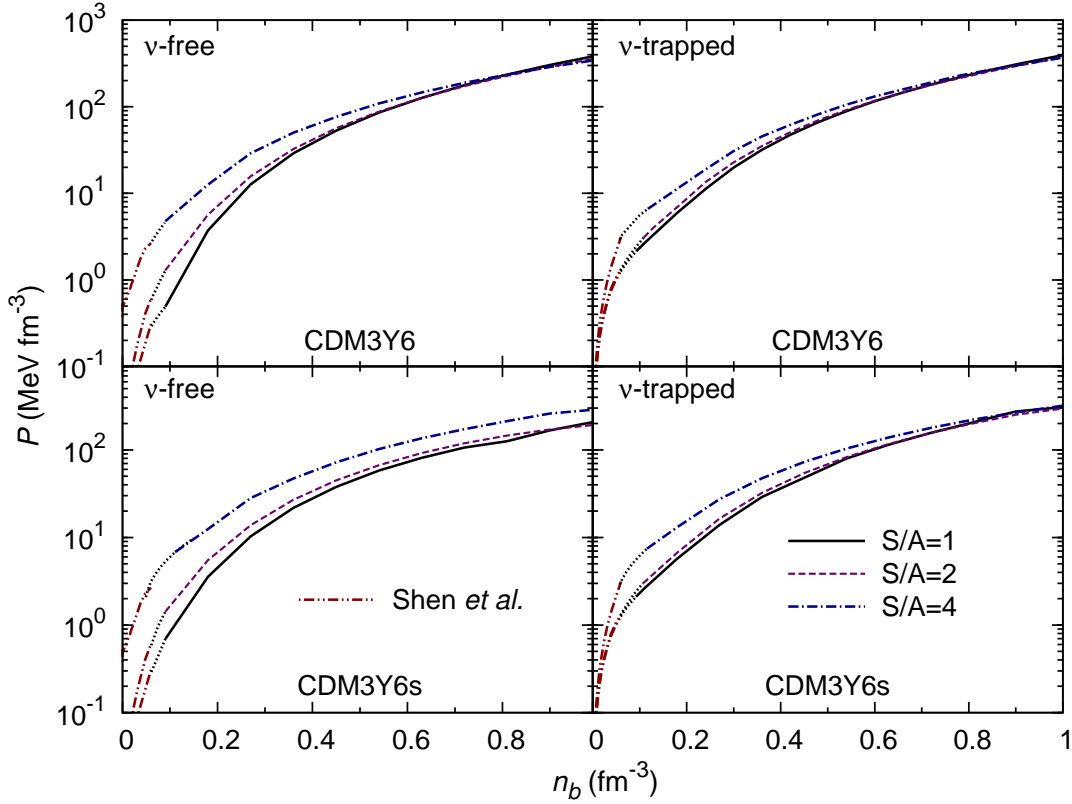} \vspace*{-0.5cm}
 \caption{Pressure (\ref{ek15}) of the isentropic $\nu$-free (left panel) and 
$\nu$-trapped (right panel) $\beta$-stable PNS matter at different baryon densities 
$n_b$ and entropy per baryon $S/A=1,2$ and 4. The EOS of the PNS crust is given 
by the RMF calculation by Shen {\it et al.} \cite{Shen,Shen11}, and the EOS of 
the uniform PNS core is given by the HF calculation using the CDM3Y6 interaction 
\cite{Loa15} (upper panel) and its soft CDM3Y6s version \cite{Loa11} (lower panel). 
The transition region matching the PNS crust with the uniform core is shown as 
the dotted lines.} \label{f11}
\end{figure}
To study the $\beta$-stable hydrostatic configuration of PNS, one needs to keep
the $\beta$ equilibrium between $n,p$ and $e,\ \mu,\ \nu$, and to combine 
the EOS of uniform matter in the core with an EOS of the hot PNS crust. 
We have considered, therefore, the EOS given by the RMF calculation by Shen 
{\it et al.} \cite{Shen} (the improved version of 2011 \cite{Shen11})
for the PNS crust. The EOS of hot inhomogeneous PNS crust given by the RMF 
calculation was used for the baryon number densities up to the edge density 
$n_{\rm edge}\approx 0.05$ fm$^{-3}$, and the EOS of the uniform PNS matter given
by our HF calculation has been used for $n_b\gtrsim 0.08$ fm$^{-3}$ following 
a brief transition region $ 0.05\lesssim n_b\lesssim 0.08$ fm$^{-3}$ shown 
as dotted lines in Fig.~\ref{f11}. The energy and pressure in the transition 
region were determined by a spline procedure to ensure their smooth and continuous 
behavior from the inner crust to the uniform core. The transition density assumed 
in the present work is rather close to that found recently in a dynamical mean-field study 
of hot neutron star matter using the momentum-dependent effective interaction \cite{Xu10}. 

The uniform PNS core is assumed to be homogeneous matter of neutrons, protons, 
electrons, neutrinos, and muons ($\mu^-$ appear at $n_b$ above the muon threshold 
density, where electron chemical potential $\mu_e>m_\mu c^2\approx 105.6$ MeV). 
In the present work, we have considered two different scenarios for neutrinos in 
the PNS matter: $\nu$-free (neutrinos are absent) and $\nu$-trapped (the PNS matter 
is opaque \cite{Bur86} and neutrinos are trapped inside the PNS). Thus, the total free 
energy density $F$ of PNS matter is determined at the given baryon number 
density $n_b$ and temperature $T$ as
\begin{eqnarray}
F(T,n_b)&=&F_b(T,n_b)+F_l(T,n_b), \label{ek11} \\
\ {\rm where}\ F_b(T,n_b)&=& E_b(T,n_b)-TS_b(T,n_b), \label{ek11b} \\
E_b(T,n_b)&=& E(T,n_b)+n_nm_nc^2+n_pm_pc^2, \label{ek11n} \\
\ {\rm and}\  F_l(T,n_b)&=&\sum_{i=e,\mu,\nu}[E_i(T,n_b,n_i)
 -TS_i(T,n_b,n_i)].  \label{ek11l} 
\end{eqnarray}
Here $E(T,n_b)$ is the HF energy density of baryons (\ref{ek1}) and 
$S_b(T,n_b)$ is the corresponding baryon entropy density (\ref{ek6}). The summation 
in (\ref{ek11l}) is done over all leptons under consideration, with the energy 
$E_i$ and entropy $S_i$ of the $i$-kind lepton evaluated in the relativistic 
(non-interacting) Fermi gas model \cite{Shapiro}. The EOS of PNS matter is 
usually discussed in terms of the total free energy per baryon and total entropy 
per baryon determined as
\begin{eqnarray}
\frac{F(n_b,T)}{A}&\equiv& [F_b(n_b,T)+F_l(n_b,T)]/n_b \label {ek11k} \\
 \frac{S(n_b,T)}{A}&\equiv& [S_b(n_b,T)+\sum_{i=e,\mu,\nu}S_i(n_b,n_i,T) ]/n_b. 
 \label {ek11m}
\end{eqnarray}
The chemical potentials of the constituent particles of PNS matter are 
constrained by the condition of $\beta$-equilibrium
\begin{equation}
\mu_n-\mu_p=\mu_e-\mu_{\nu_e}=\mu_\mu-\mu_{\nu_\mu}. \label {ek11t}
\end{equation}  
With the fraction of the $k$-kind constituent particle determined at the  
baryon number density $n_b$ as $x_k=n_k/n_b$, the charge neutrality condition 
of $\beta$-stable PNS matter implies
\begin{equation}
 x_p+\sum_{i=e,\mu,\nu} q_i x_i=0, \label {ek12}
\end{equation}
where $q_i$ is the charge of the $i$-kind lepton. The conservation of the weak
charge leads to the conservation of the lepton fractions determined as 
\begin{equation}
 Y_l=x_l-x_{\bar{l}}+x_{\nu_l}-x_{\bar{\nu}_l},\ \ l=e,\ \mu. \label {ek13}
\end{equation}
Because of dynamical neutrino trapping during the core collapse \cite{Bet90}, 
neutrinos are \emph{trapped} during the first instants of the PNS \cite{Bur86,Pra97}.
At the onset of neutrino trapping occuring at $\rho_b\sim 10^{11}$g/cm${^3}$ 
\cite{Kotake}, the electron lepton fraction is expected to be $Y_e\approx 0.4$ 
\cite{Pra97,Bur10,Li10}. When the electron lepton fraction is fixed, the conversion 
of electrons to muons at high densities is unlikely and the muon lepton fraction can be 
assumed $Y_\mu\approx 0$ \cite{Pra97,Bur10}. These assumptions for the $\nu$-trapped 
PNS matter, together with a chosen EOS for the baryon matter, are the constraints 
used to determine the $npe\mu\nu$ composition of the $\beta$-stable, $\nu$-trapped 
PNS matter. In the $\nu$-free case, these conditions are simplified by setting all 
neutrino fractions to zero. We note that at very low baryon densities 
of $n_b<10^{-5}$ fm$^{-3}$ neutrinos are no more trapped inside the dilute PNS 
atmosphere and become almost free \cite{Fish09}. We have chosen, therefore, also 
a ``cut-off"' procedure to impose the electron fraction $x_e\approx 0.4$ at 
$n_b\lesssim 10^{-5}$ fm$^{-3}$and the dilute PNS atmosphere is treated as $\nu$-free.

\begin{figure}\hspace*{-1.0cm}
\includegraphics[width=1.0\textwidth] {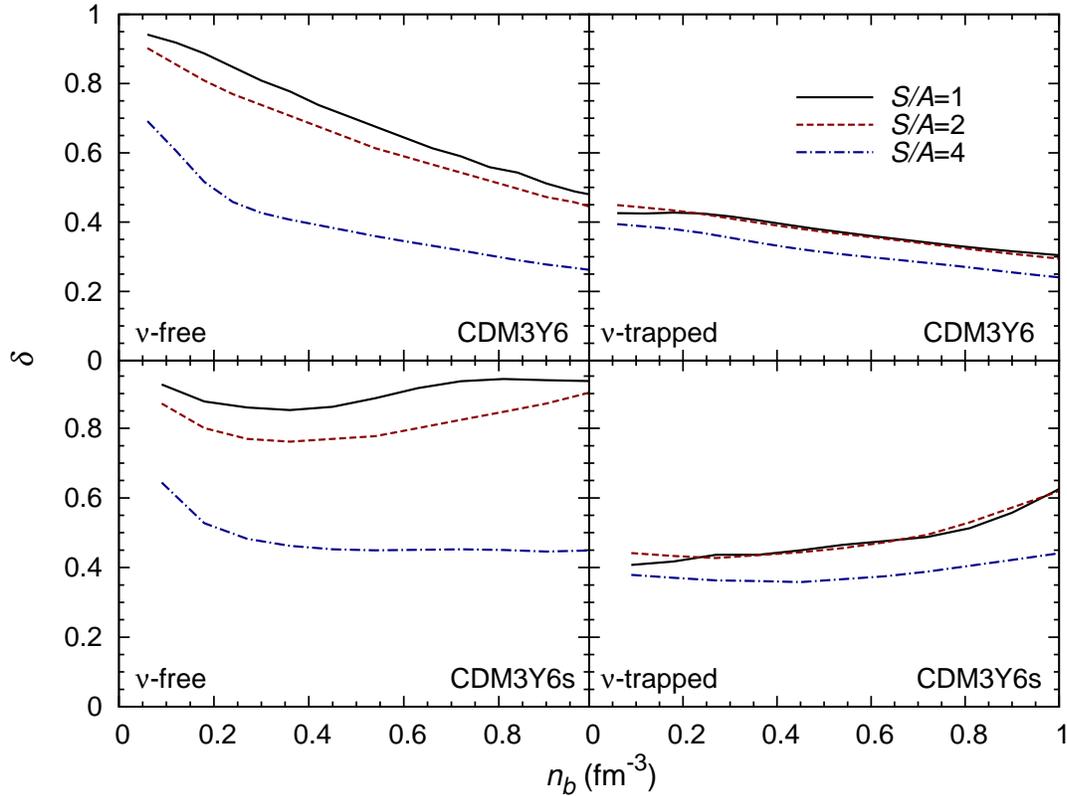} \vspace*{-1cm}
 \caption{Neutron-proton asymmetry $\delta$ of the $\nu$-free (left panels) and 
$\nu$-trapped (right panels) $\beta$-stable PNS matter at different baryon number 
densities $n_b$ and entropy per baryon $S/A=1,2$ and 4. The EOS of the homogeneous 
PNS core is given by the HF calculation using the CDM3Y6 interaction \cite{Loa15} 
and its soft CDM3Y6s version \cite{Loa11}.} \label{f12}
\end{figure}
At zero temperature, a simple relation linking $\mu_n-\mu_p$ and the nuclear 
symmetry energy based on the parabolic approximation (\ref{ek9})
is often used to determine the proton, electron and muon fractions 
of $\beta$-stable matter \cite{Loa11}. Because the parabolic approximation
becomes poorer with the increasing temperature as shown in Sect.~\ref{sec1}, we 
avoid using it in the present HF calculation. Thus, at each temperature $T$ the 
EOS associated with a given effective NN interaction is first constructed in a 
dense grid of the neutron-proton asymmetries $\delta$ and baryon number densities 
$n_b$. Then, at the given temperature and baryon density, the actual asymmetry 
parameter $\delta$ is obtained by ensuring the conditions of $\beta$-equilibrium 
(\ref{ek11}), charge neutrality (\ref{ek12}) and lepton number conservation 
(\ref{ek13}). At each step of this variation procedure, the chemical potentials 
of constituent particles are determined by normalizing the Fermi-Dirac momentum 
distribution to the corresponding number density at the given $T,\ n_b,$ and 
$\delta$. With the obtained PNS composition, the temperature dependent EOS 
of $\beta$-stable $npe\mu\nu$ matter is fully given by the total free energy 
density $F(T,n_b)$ determined by Eq.~(\ref{ek11}) and total pressure $P(T,n_b)$ 
determined as 
\begin{equation}
 P(T,n_b) = n_b^2{\frac{\partial}{{\partial n_b}}}\left[\frac{F_b(T,n_b)}
{n_b}\right]+\sum_{i=e,\mu,\nu} P_i (T,n_b,n_i). \label{ek15}
\end{equation}
Like the lepton internal energy $E_i$, the pressure of the $i$-kind lepton $P_i$ 
is also determined by the relativistic free Fermi gas model \cite{Shapiro}.
The total entropy density $S(T,n_b)$ at each baryon density was also determined in 
a dense grid of different temperatures $T$, so that the isentropic behavior 
of all considered thermodynamic quantities can be easily interpolated at
different baryon densities. For example, the total pressure $P$ representing 
the EOS of the isentropic, $\nu$-free and $\nu$-trapped $\beta$-stable matter 
is shown in Fig.~\ref{f11}. One can see that the difference between the soft- 
and stiff symmetry energy scenarios is quite significant in the $\nu$-free case,
while it becomes somewhat weaker for the $\nu$-trapped PNS matter. 

\begin{figure}
\includegraphics[width=1.0\textwidth]{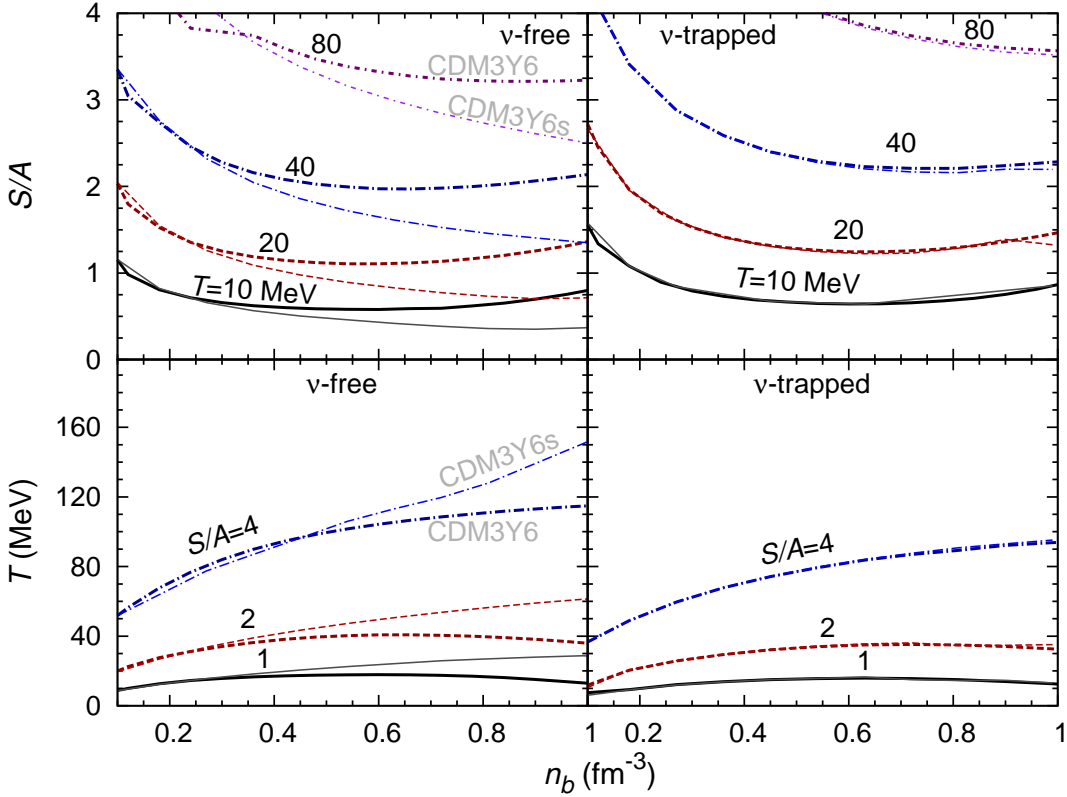}\vspace*{-0.5cm}
 \caption{(Color online) Entropy per baryon (upper panel) and temperature 
(lower panel) as function of baryon number density $n_b$ of the $\beta$-stable PNS 
matter given by the CDM3Y6 interaction \cite{Kho97,Loa15} (thick lines) and its 
soft CDM3Y6s version \cite{Loa11} (thin lines) in the $\nu$-free (left panel) 
and $\nu$-trapped (right panel) cases.} \label{f13}
\end{figure}
At variance with the HF calculation of NM, the neutron-proton asymmetry $\delta$ 
in the $\beta$-stable PNS matter becomes a thermodynamic quantity, determined 
consistently by the (equilibrated) neutron and proton fractions at temperature 
$T$ and baryon number density $n_b$. The $\delta$ values given by the HF calculation 
using the CDM3Y6 and CDM3Y6s interactions for the isentropic $\beta$-stable 
$npe\mu\nu$ matter at the total entropy per baryon $S/A=1,2$ and 4 are shown in 
Fig.~\ref{f12}. One can see that for both the stiff- and soft symmetry-energy 
scenarios, the hot PNS matter with high entropy of $S/A=4$ becomes more proton 
rich compared to that at the lower temperature or entropy. In the $\nu$-free 
case (see left panel of Fig.~\ref{f12}), the $\delta$ value given by the stiff 
CDM3Y6 interaction at $S/A=1,2$ decreases gradually from about 0.9 at 
low densities to around 0.5 at $n_b=1$ fm$^{-3}$. The same trend was found at 
$S/A=4$, but the matter is more symmetric or more proton rich. The behavior of 
$\delta$ value given by the soft CDM3Y6s interaction at entropy $S/A=1,2$ is 
similar to that found earlier \cite{Loa11} for the $\beta$-stable matter at 
zero temperature, with the core matter remaining very neutron rich over the 
whole density range. As temperature increases to the thermal equilibrium at $S/A=4$, 
the difference between the two scenarios for the symmetry energy becomes less 
sizable and $\delta$ value decreases to around 0.25-0.45 at the highest baryon 
density. The neutrino trapping was found to make the difference between the 
stiff and soft symmetry-energy scenarios less significant (see right panel of 
Fig.~\ref{f12}), with the matter becoming more proton rich over a wide range 
of baryon densities. Such an enhancement of the proton fraction in PNS matter 
by neutrino trapping was also found in the BHF calculation by Vida\~na {\it et al.} 
\cite{Vid03} that includes hyperons. The suppression of the impact of the 
nuclear symmetry energy in the presence of trapped neutrinos is mainly explained 
by the fact that the PNS matter is more symmetric and weak processes, like 
$p+e^- \leftrightarrow n+\nu_e$, can proceed in both directions. It is clearly 
seen in Fig.~\ref{f12} that the more symmetric matter the less impact of the
symmetry energy. This effect is also illustrated in the density profiles of entropy 
and temperature of the $\beta$-stable PNS matter given by the CDM3Y6 and CDM3Y6s 
interactions shown in Fig.~\ref{f13}. One can see on the left panel of Fig.~\ref{f13} 
that in the $\nu$-free case, the difference between the stiff and soft symmetry-energy 
scenarios is strongest at high baryon densities so that the center of PNS becomes 
much hotter when the symmetry energy is soft. Such an impact to the density profiles 
of entropy and temperature by the symmetry energy is, however, suppressed in the 
presence of trapped neutrinos (see right panel of Fig.~\ref{f13}). 
\begin{figure}
\includegraphics[width=0.9\textwidth]{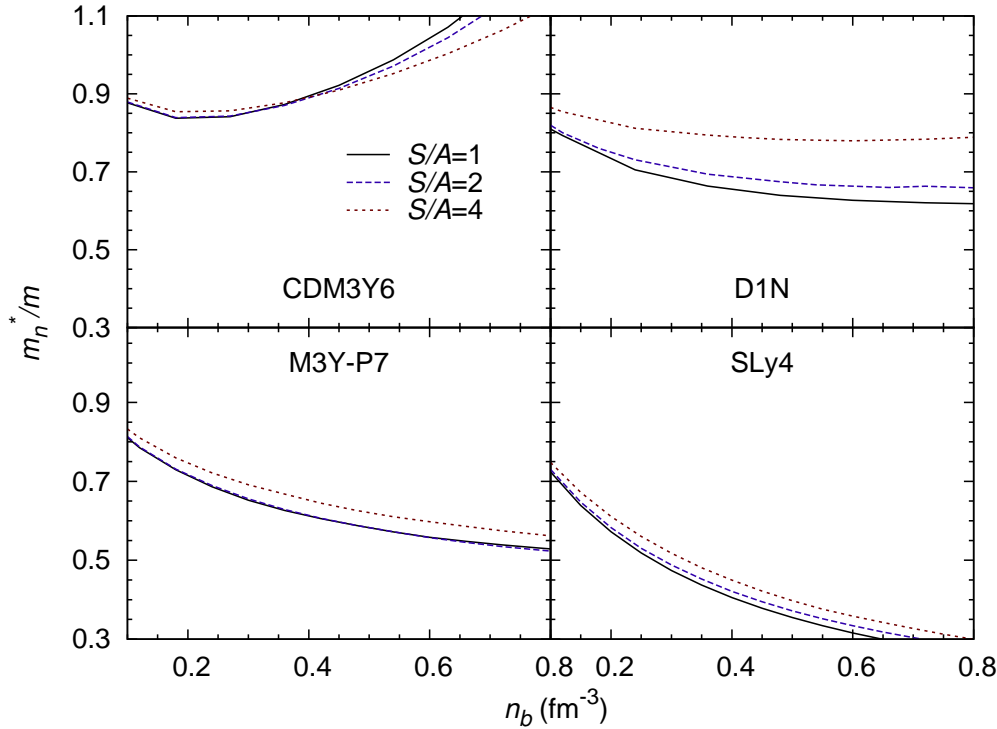}\vspace*{-0.5cm}
 \caption{(Color online) Density profile of neutron effective mass in 
the $\nu$-free and $\beta$-stable PNS matter at entropy per baryon 
$S/A=1,2$ and 4, given by the HF calculation using the CDM3Y6 \cite{Loa15} and 
M3Y-P7 \cite{Na13} interactions (left panel), the D1N version of Gogny 
interaction \cite{Ch08} (right panel) and SLy4 version \cite{Ch98} of Skyrme 
interaction (right panel).} \label{f4n}
\end{figure}
\begin{figure}
\includegraphics[width=0.9\textwidth]{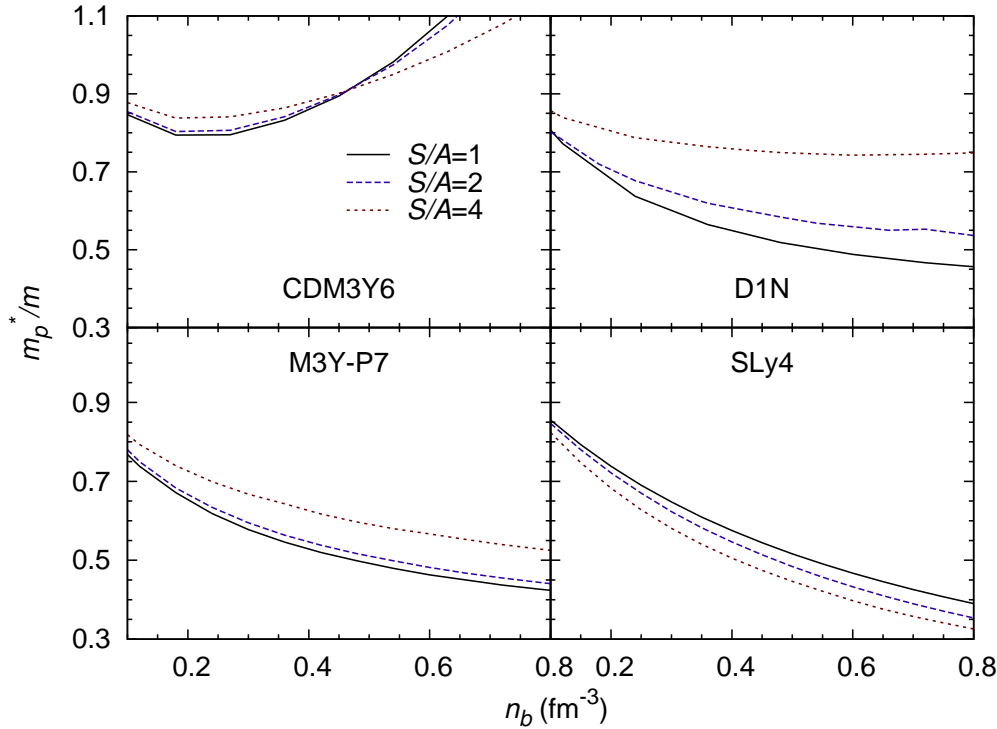}\vspace*{-0.5cm}
\caption{(Color online) The same as Fig.~\ref{f4n} but for proton
effective mass.} \label{f5n}
\end{figure}

Beside the symmetry energy, the impact of nucleon effective mass to the 
thermal properties of NM was found quite significant in the previous section,
and it is of interest to explore the effect of $m^*_\tau$ in the $\beta$-stable 
PNS matter at different temperatures and entropy.  
\begin{figure}
\includegraphics[width=0.9\textwidth]{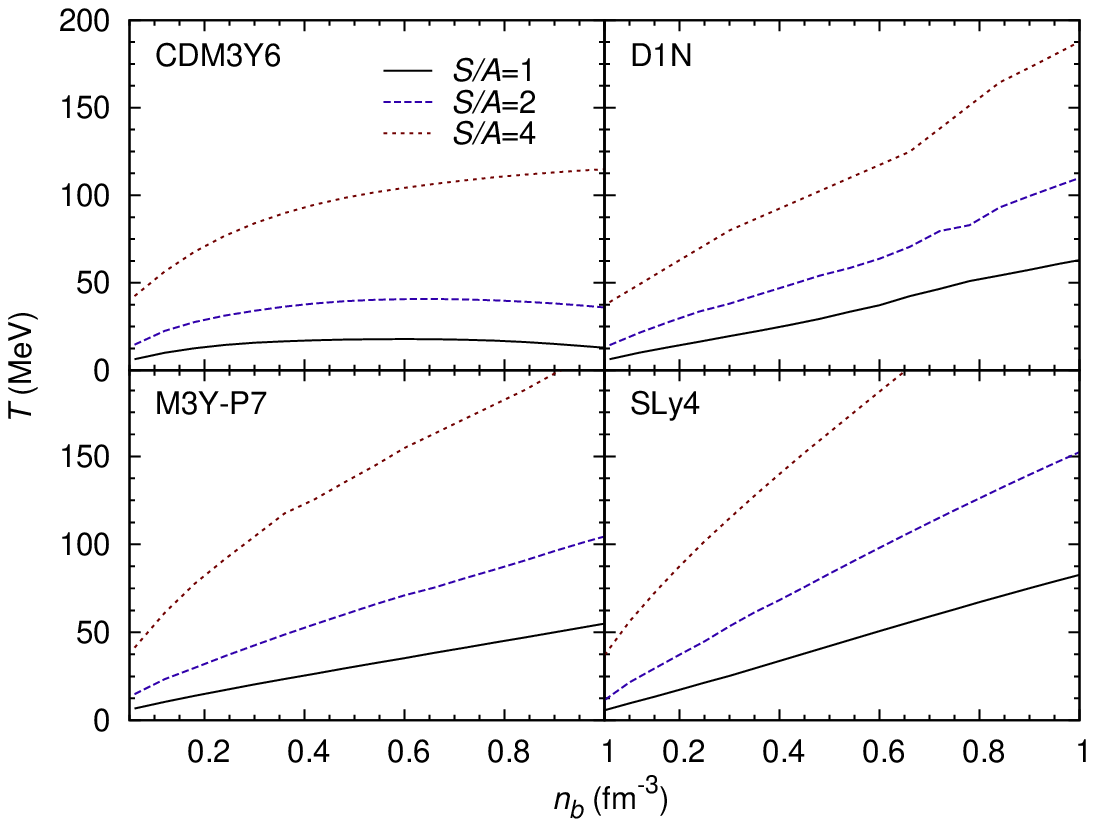}\vspace*{-0.5cm}
 \caption{(Color online) Density profile of temperature in the $\nu$-free 
and $\beta$-stable PNS matter at entropy per baryon $S/A=1,2$ and 4,
deduced from the HF results obtained with the same density dependent NN 
interactions as those considered in Fig.~\ref{f4n}.} \label{f6n}
\end{figure}
The density profiles of neutron and proton effective masses in the $\nu$-free and 
$\beta$-stable PNS matter at entropy per baryon $S/A=1,2$ and 4, obtained with
different density dependent NN interactions are shown in Figs.~\ref{f4n} and
\ref{f5n}. One can see that the density dependence of nucleon effective mass
in the hot $\beta$-stable PNS matter is quite similar to that at zero temperature
shown in Figs.~\ref{f1n} and \ref{f2n}. While nucleon effective mass 
given by the CDM3Y6 interaction saturates at $n_b\approx 0.2$ fm$^{-3}$ and rises 
up to above unity at high baryon densities (in agreement with the BHF calculation 
\cite{Bal14} that includes 3-body forces), the $m^*_\tau$ values given by other 
interactions steadily decrease with the increasing baryon density. The fall of 
nucleon effective mass given by the M3Y-P7 and Sly4 interactions is very drastic 
at high densities and it should result, therefore, on very high temperature in the 
center of PNS. The density profiles of temperature in the $\nu$-free and $\beta$-stable 
PNS matter at entropy per baryon $S/A=1,2$ and 4, given by the same density dependent 
NN interactions are shown in Fig.~\ref{f6n}. It can be seen that the rise of temperature 
with the increasing baryon density given by the M3Y-P7 and Sly4 interactions is indeed 
very stiff, to $T$ above 200 MeV at high densities. We will see below that such 
different behaviors of nucleon effective mass also influence strongly the hydrostatic 
configuration of hot PNS.      
 
\begin{figure}
\includegraphics[width=1.0\textwidth] {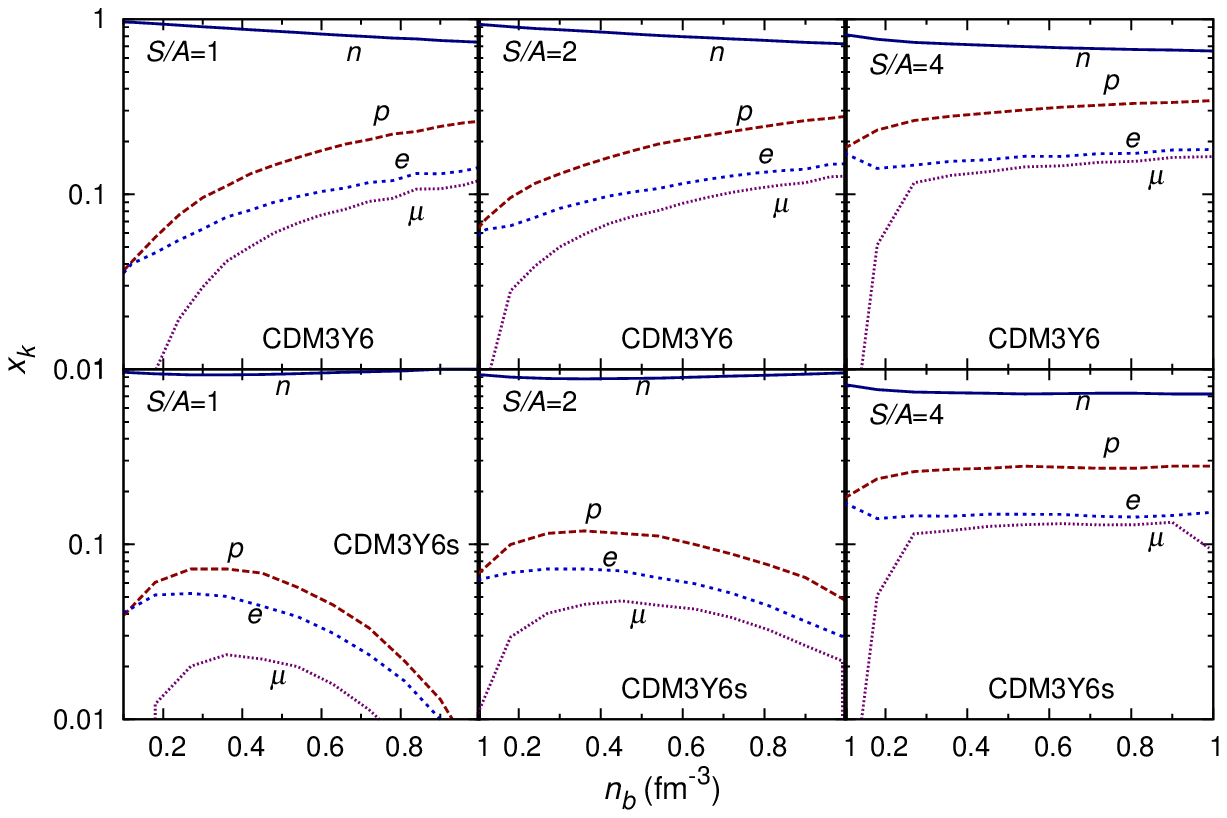} \vspace*{-1.5cm} 
\caption{(Color online) Particle fractions as function of baryon number 
density $n_b$ in the $\nu$-free and $\beta$-stable PNS matter at entropy per 
baryon $S/A=1,2$ and 4, given by the CDM3Y6 interaction 
\cite{Loa15} (upper panel) and its soft CDM3Y6s version \cite{Loa11}
(lower panel).} \label{f14}
\end{figure}
\begin{figure}
\includegraphics[width=1.0\textwidth] {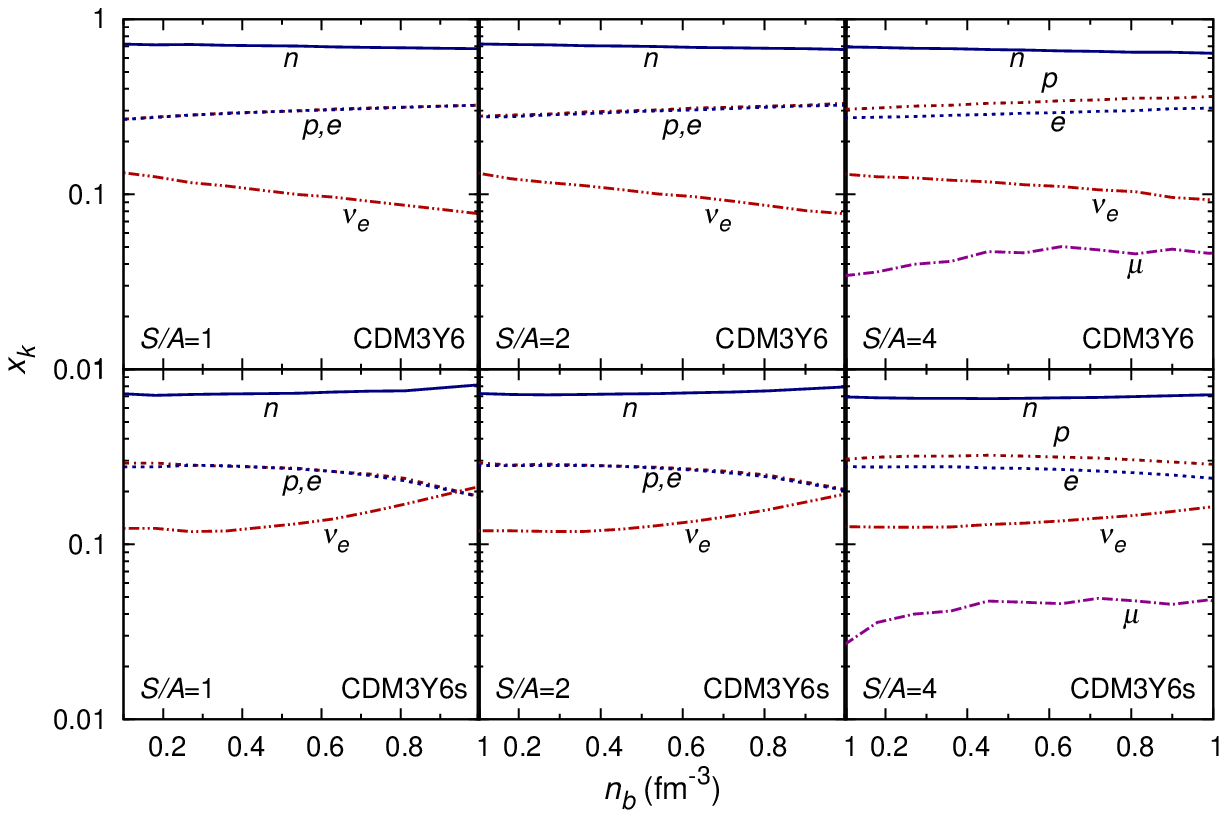} \vspace*{-1.5cm} 
 \caption{The same as Fig.~\ref{f14} but for the $\nu$-trapped,
  $\beta$-stable matter of the PNS.} \label{f15}
\end{figure}
\begin{figure}
\includegraphics[width=1.0\textwidth] {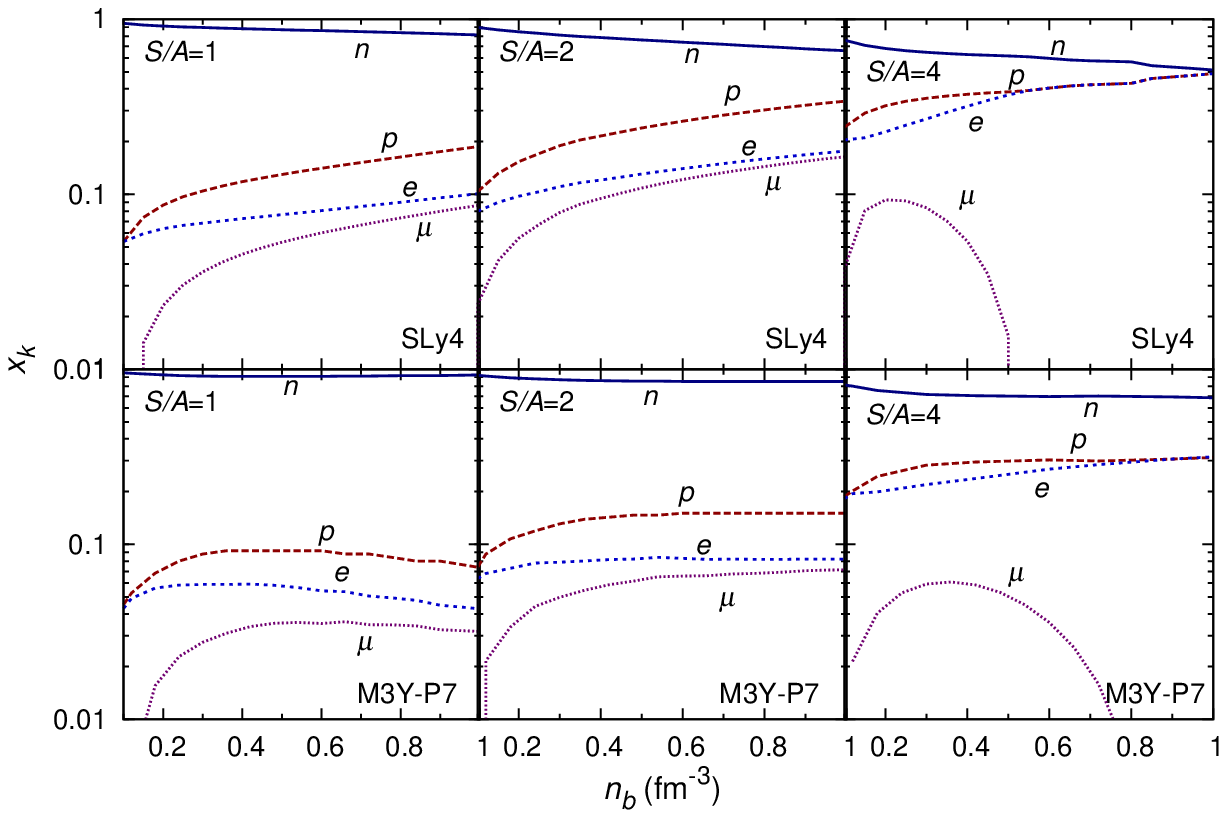} \vspace*{-1.5cm} 
\caption{(Color online) The same as Fig.~\ref{f14}, but given by the SLy4 version 
\cite{Ch98} of Skyrme interaction (upper panel) and M3Y-P7 interaction 
parametrized by Nakada \cite{Na13} (lower panel).} \label{f16}
\end{figure}
\begin{figure}
\includegraphics[width=1.0\textwidth] {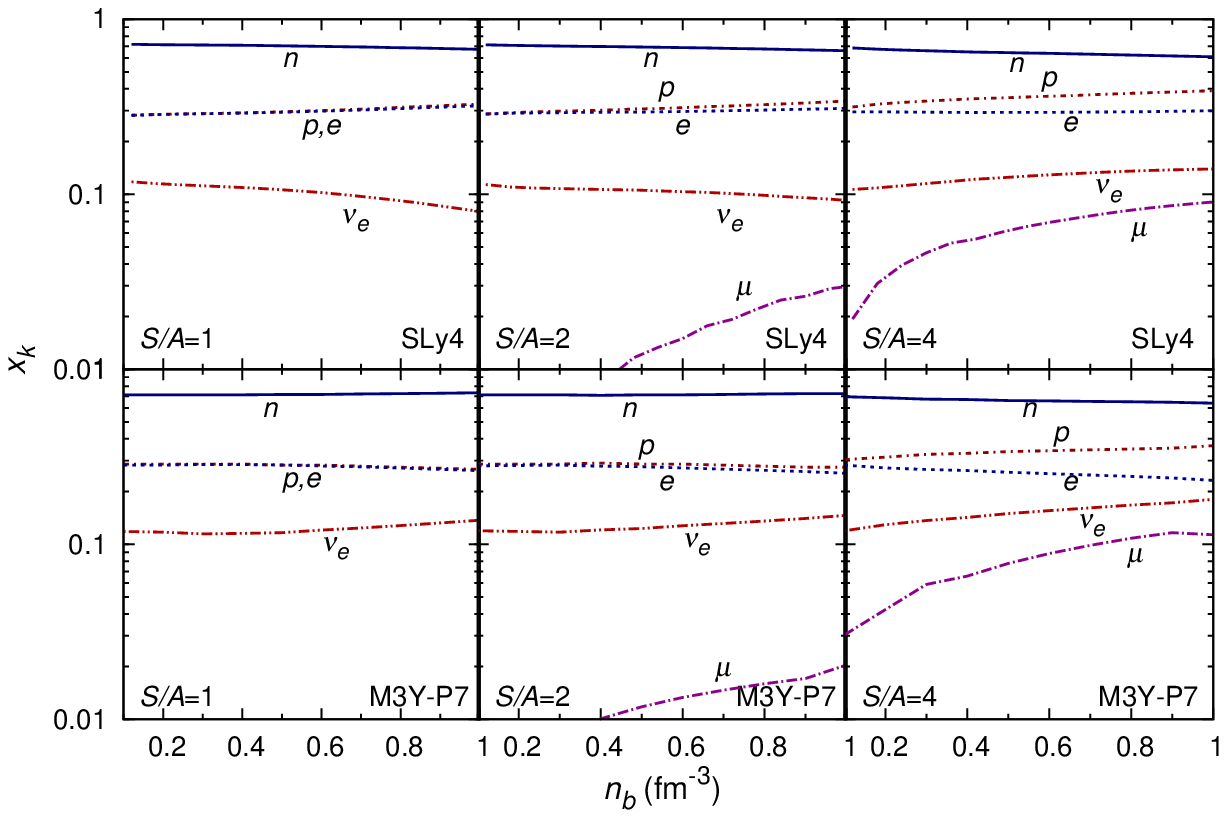} \vspace*{-1.5cm} 
 \caption{The same as Fig.~\ref{f16} but for the $\nu$-trapped,
  $\beta$-stable PNS matter.} \label{f17}
\end{figure}

Based on different EOS's given by different density dependent NN interactions, 
the composition of the $\beta$-stable PNS in terms of the constituent-particle 
fractions at entropy per baryon $S/A=1,2$ and 4 is quite different (see 
Figs.~\ref{f14}-\ref{f17}). The results shown in Figs.~\ref{f14} and 
\ref{f16} were obtained for the $\nu$-free PNS matter that corresponds the late 
stage of the evolution of PNS when most of neutrinos have escaped from the core. 
In this case the difference between the stiff- and soft symmetry-energy scenarios 
is quite significant as discussed above for the neutron-proton asymmetry.  
A comparison with the BHF results obtained for the $\beta$-stable PNS matter 
at entropy $S/A=1,2$ (see upper panels of Figs.~\ref{f14} and \ref{f16}, 
and Fig.~4 of Ref.~\cite{Bur10}) shows that the asy-stiff interactions (CDM3Y3, 
CDM3Y6, and SLy4) give the results rather close to the BHF results, while the 
results given by the asy-soft interactions (CDM3Y3s, CDM3Y6s, and M3Y-P7) are 
very different from the BHF results. The results shown in Figs.~\ref{f15} and 
\ref{f17} were obtained for the $\nu$-trapped PNS matter with the electron lepton 
fraction $Y_e\approx 0.4$, as expected during the first stage of the core-collapse 
supernovae \cite{Pra97,Bur10,Li10}. In such a scenario, the electron conversion 
to muon is unlikely at low temperatures or entropy, and the muon fraction was found 
much suppressed at $S/A=1$ and 2. At high temperatures of the PNS matter at $S/A=4$, 
the electron chemical potential $\mu_e$ increases and exceeds the muon threshold, 
and muons appear again. 

In both cases, the effect of the symmetry energy is clearly demonstrated: 
at high baryon densities, the asy-stiff interactions always give higher electron 
and proton fractions compared with those given by the asy-soft interactions. 
Such an impact of the symmetry energy is more pronounced in the $\nu$-free case, 
where the electron fraction $x_e$ predicted by the asy-soft models reduces 
significantly to below 0.1 at high baryon densities. As discussed in Ref.~\cite{Loa11}, 
the difference in the electron fraction caused by different slopes of the 
symmetry energy at high baryon densities has a drastic effect on the 
$\nu$-emission process during the cooling of neutron star. Namely, in the soft 
symmetry-energy scenario, the direct Urca process is strongly quenched because of
$Y_e < 11\%$ \cite{Loa11,Lat91}, while it is well allowed in the stiff symmetry-energy 
scenario. The difference in the electron fraction caused by different slopes 
of the symmetry energy is less pronounced in very hot PNS matter at $S/A=4$. 
Such state of hot PNS matter has been shown to occur at the onset of collapse 
of a very massive (40 $M_\odot$) progenitor to black hole \cite{Hem12,Ste13}.    

It is interesting to note the difference between results obtained in the 
$\nu$-trapped and $\nu$-free cases. Although $\beta$-equilibrium implies the 
weak processes $l+p \leftrightarrow n+ \nu_l$, in the $\nu$-free PNS matter 
most of neutrinos have escaped and cannot contribute to these processes, 
and they proceed preferably in one direction (from left to right). As a result, 
the PNS matter undergoes the neutronization process \cite{Bet90, Reddy00}. 
Therefore, the $\nu$-free PNS matter is always more neutron-rich than the 
$\nu$-trapped matter as shown above in Fig.~\ref{f12}. Finally we discuss also 
the effect of increasing temperature (or equivalently increasing entropy) on the 
particle fractions shown in Figs.~\ref{f14}-\ref{f17}. In general, the increase 
of temperature tends to reduce the difference between the neutron and proton 
fractions ($n_n$ decreases while $n_p$ increases  with $T$). In a similar way, 
the difference between $e$ and $\mu$ becomes also less pronounced with the 
increasing temperature ($n_e$ decreases and $n_\mu$ increases with $T$). 
Such a phenomenon is expected to occur between particles of similar nature, 
and the hot matter tends to quench the difference in their masses. Although 
the impact of the symmetry energy in the presence of trapped neutrinos was 
shown to be less significant with increasing entropy, we found that the symmetry 
energy still has some effect on the neutrino fraction in the high-density 
$\nu$-trapped PNS matter at $S/A=4$. Even with a fixed electron lepton fraction 
$Y_e=0.4$, the neutrino fraction $x_\nu<10\%$ was found with the asy-stiff CDM3Y6 
interaction as $n_b$ approaches 1 fm$^{-3}$, while $x_\nu$ close to 20\% was 
found with the soft CDM3Y6s interaction (see right panel of Fig.~\ref{f15}). 
The result given by the stiff CDM3Y6 interaction seems to agree well with those 
of the recent hydrodynamic simulation of black hole formation in the failed 
core-collapse supernova \cite{Hem12}, which suggested that the thermodynamic 
condition of hot $\beta$-stable PNS matter at the onset of collapse of a massive
(40 $M_\odot$) progenitor to black hole can be approximated by a constant entropy 
per baryon $S/A\approx 4$, with a \emph{negligible} contribution of neutrinos 
($Y_{\nu_e}+Y_{\bar\nu_e}<0.05$ at $0.2<Y_e<0.3$ as shown in Fig.~16 
of Ref.~\cite{Hem12}).     

\section{$\beta$-stable configuration of hot protoneutron star}\label{sec3}
\begin{figure}
\includegraphics[width=1.0\textwidth] {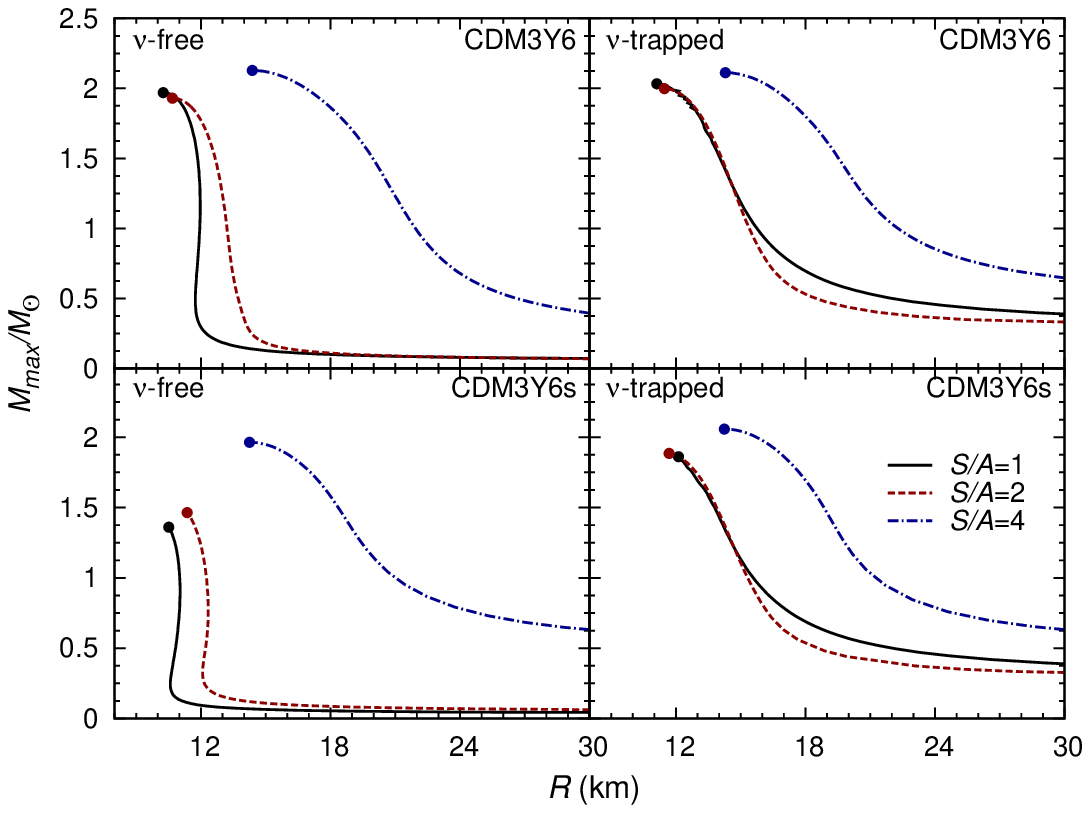} \vspace*{-1.0cm} 
\caption{(Color online) Gravitational mass (in unit of solar mass $M_\odot$) of the 
$\beta$-stable, $\nu$-free (left panel) and $\nu$-trapped (right panel) PNS at  
entropy $S/A=1,2$ and 4 as function of the radius (in km), based on the EOS of the 
homogeneous PNS core given by the CDM3Y6 interaction \cite{Loa15} (upper panel) and 
its soft CDM3Y6s version \cite{Loa11} (lower panel). The circle at the end of each curve 
indicates the last stable configuration.} \label{f18}
\end{figure}
\begin{figure}
\includegraphics[width=1.0\textwidth] {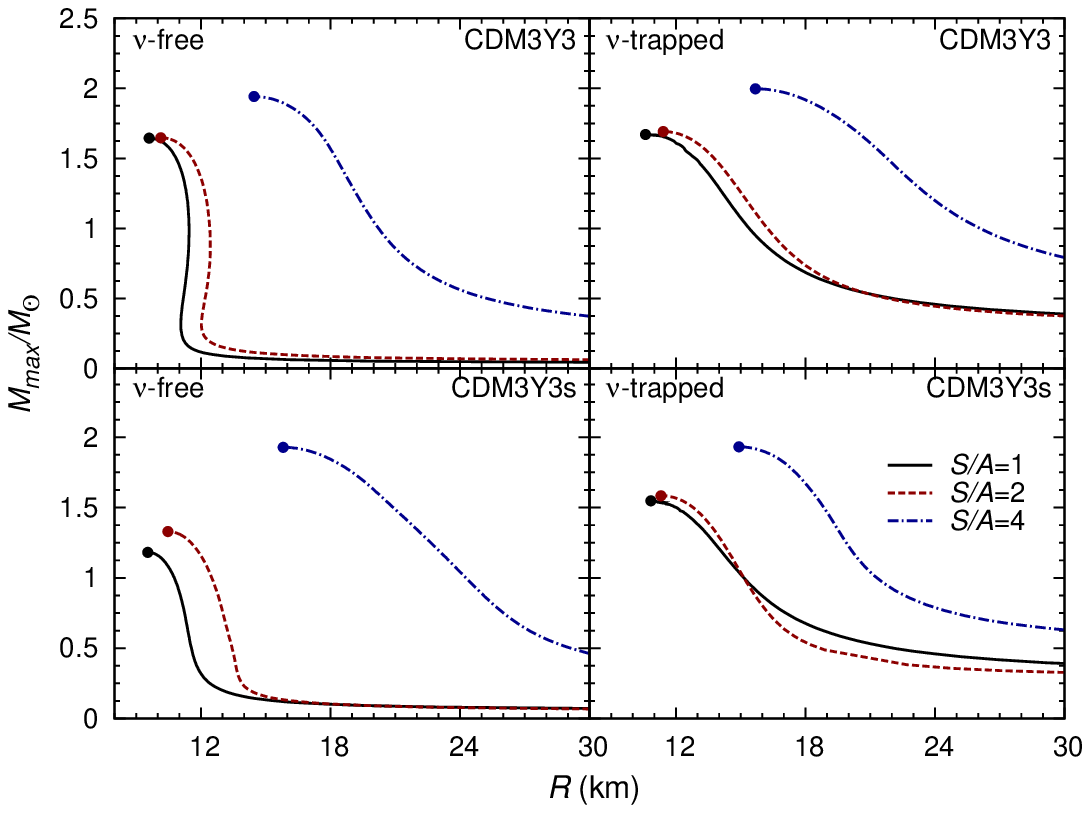} \vspace*{-1.0cm} 
\caption{The same as Fig.~\ref{f18} but given by the CDM3Y3 interaction 
\cite{Loa15} (upper panel) and its soft CDM3Y3s version \cite{Loa11} (lower panel).}
\label{f19}
\end{figure}
\begin{figure}
\includegraphics[width=1.0\textwidth] {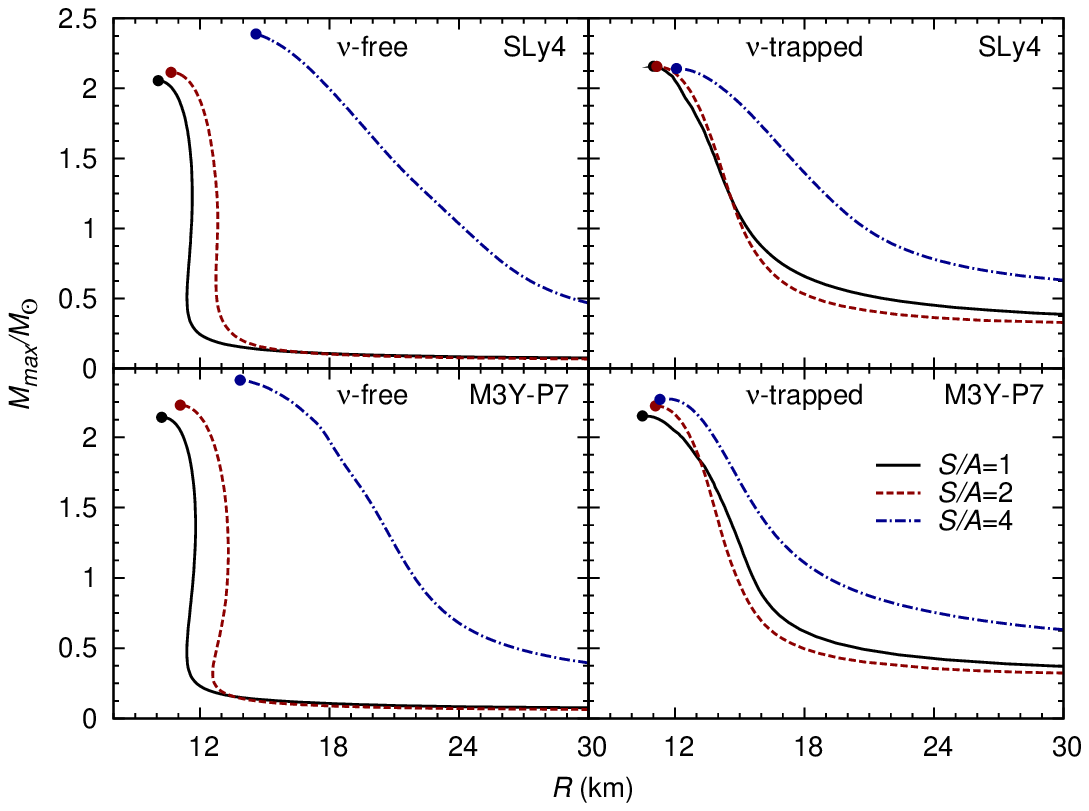} \vspace*{-1.0cm} 
\caption{The same as Fig.~\ref{f18} but given by the SLy4 version of Skyrme 
interaction \cite{Ch98} (upper panel) and M3Y-P7 interaction 
parametrized by Nakada \cite{Na13} (lower panel).}\label{f20}
\end{figure}

We used the total internal energy $E$ and pressure $P$ inside PNS at entropy per 
baryon $S/A=1,2$ and 4 given by the different EOS's discussed above in Sec.~\ref{sec2} 
as inputs for the Tolman-Oppenheimer-Volkov (TOV) equations \cite{Shapiro}. The 
$\beta$-stable hydrostatic configuration of hot PNS in different scenarios given by 
the solutions of the TOV equations are presented in Tables~\ref{t2} and \ref{t3}, 
and Figs.~\ref{f18}-\ref{f20}. For comparison, we have also presented in these 
tables the $\beta$-stable configuration of cold ($\nu$-free) neutron star, given by 
the solutions of the TOV equations obtained with the same EOS's \cite{Loa11}.
    
\begin{table}
\setlength{\tabcolsep}{.3cm}
\renewcommand{\arraystretch}{0.7}
\caption{Properties of the $\nu$-free and $\nu$-trapped, $\beta$-stable PNS at 
entropy per baryon $S/A=1,2$ and 4, given by the solutions of the TOV equations 
using the EOS's based on the CDM3Y3, CDM3Y6 interactions \cite{Loa15} and their
soft CDM3Y3s, CDM3Y6s versions \cite{Loa11}. $M_{\rm max}$ and $R_{\rm max}$ 
are the maximum gravitational mass and radius; $n_c,\ \rho_c,\ P_c$, and $T_c$ 
are the baryon number density, mass density, total pressure, and temperature
in the center of PNS. $T_s$ is temperature of the outer core of PNS, at baryon 
density $\rho_s\approx 0.63\times 10^{15}$ g/cm$^3$. Results at $S/A=0$ represent 
the stable configuration of cold ($\nu$-free) neutron star \cite{Loa11}.} \label{t2}
\vspace{0.3cm}
\centering
\begin{tabular}{ccccccccc} \hline
EOS & $S/A$ & $M_{\rm max}$&  $R_{\rm max}$ & $n_c$ & $\rho_c$ &$P_c$ & $T_c$ & $T_s$ \\ 
 & & ($M_\odot$) & (km) & (fm$^{-3}$) & ($10^{15}$g/cm$^3$) & (MeV~fm$^{-3}$) 
 & (MeV) & (MeV) \\ \hline        
CDM3Y3  & 0  &  1.59 & 9.70 & 1.44  & 3.16 & 494.4 & 0.0 & \\  
($\nu$-free) & 1 &  1.64  & 9.60  & 1.53  & 3.16 & 538.9 & 19.7 & 17.6 \\  
   & 2  &   1.65  & 10.14 & 1.40  & 2.90 & 446.8 & 49.4 & 37.9 \\  
         &      4 &   1.94  & 14.67 & 0.78  & 1.58 & 180.5 & 113.6 & 84.6 \\ 
CDM3Y3  & 1 &  1.67  & 10.59 & 1.44  & 2.90 & 470.8 & 18.7 & 14.3 \\
($\nu$-trapped) & 2 &   1.69  & 11.40 & 1.29  & 2.59 & 373.9 & 44.7 & 30.1 \\ 
        &     4 &   1.99  & 15.67 & 0.75  & 1.45 & 163.1 & 91.4 & 66.9 \\ \hline  
CDM3Y3s  & 0 &   1.13  &  9.36 & 1.61  & 3.26 & 261.1 & 0.0 & \\  
($\nu$-free)  & 1 &   1.18  &  9.53 & 1.52  & 3.07 & 253.0 & 37.4 & 18.4 \\  
        & 2 & 1.33  & 10.47 & 1.30  & 2.66 & 239.0 & 71.4 & 39.0 \\  
        &   4 &   1.93  & 15.28 & 0.67  & 1.33 & 130.6 & 120.3 & 86.1 \\   
CDM3Y3s   &1 &  1.55  & 10.84 & 1.46  & 2.90 & 390.3 & 18.0 & 14.2 \\  
($\nu$-trapped) & 2 &   1.58  & 11.31 & 1.30  & 2.59 & 319.0 & 46.5 & 30.1 \\  
        &   4 &   1.93  & 14.91 & 0.77  & 1.50 & 161.8 & 91.5 & 69.0 \\  \hline  
CDM3Y6    &  0 &  1.95  & 10.23 & 1.20  & 2.74 & 627.3 & 0.0 & \\ 
($\nu$-free)  & 1 &  1.97  & 10.25 & 1.24  & 2.73 & 626.5 & 10.1 & 17.4 \\  
            & 2 &   1.93  & 10.68 & 1.18  & 2.59 & 524.1 & 33.7 & 37.5 \\  
            & 4 &   2.12  & 14.37 & 0.80  & 1.63 & 235.5 & 114.1 & 86.8 \\  
CDM3Y6   & 1 &   2.03  & 11.11 & 1.18  & 2.51 & 553.9 & 9.7 & 18.3 \\   
($\nu$-trapped) & 2 &   2.00  & 11.46 & 1.13  & 2.37 & 467.7 & 24.8 & 29.6 \\  
        &   4 &   2.11  & 14.29 & 0.82  & 1.68 & 246.6 & 89.7 & 66.6 \\  \hline
CDM3Y6s & 0 &  1.42  & 9.74 & 1.46  & 3.06 & 340.4 & 0.0 & \\  
($\nu$-free) & 1 &  1.65  & 9.20 & 1.50  & 3.45 & 652.7 & 17.1 & 18.3  \\   
      & 2 &   1.37  & 11.13 & 1.16  & 2.44 & 212.9 & 65.4 & 38.8 \\   
        &    4 &   1.96  & 14.24 & 0.80  & 1.67 & 206.9 & 121.5 & 84.5 \\ 
CDM3Y6s & 1 &   2.04  & 11.67 & 1.02  & 2.02 & 314.7 & 12.3 & 13.9 \\  
($\nu$-trapped) & 2 &   1.88  & 11.68 & 1.12  & 2.30 & 353.9 & 34.6 & 29.8\\ 
        &    4 &   2.06  & 14.24 & 0.84  & 1.68 & 229.4 & 91.2 & 66.7\\  \hline  
\end{tabular} 
\end{table}
\begin{table}
\setlength{\tabcolsep}{.3cm}
\renewcommand{\arraystretch}{0.7}
\caption{The same as Table~\ref{t2} but obtained with the EOS's based on 
the SLy4 version \cite{Ch98} of Skyrme interaction, M3Y-P7 
interaction parametrized by Nakada \cite{Na13}, and D1N version 
\cite{Ch08} of Gogny interaction.} \label{t3}\vspace{0.5cm}
\centering
\begin{tabular}{ccccccccc} \hline
EOS & $S/A$ & $M_{\rm max}$&  $R_{\rm max}$ & $n_c$ & $\rho_c$ &$P_c$ & $T_c$ & $T_s$ \\ 
 & & ($M_\odot$) & (km) & (fm$^{-3}$) & ($10^{15}$g/cm$^3)$ & (MeV~fm$^{-3}$) & 
 (MeV) & (MeV)  \\ \hline        
SLy4  & 0 & 2.05   & 9.96 & 1.21  & 2.86 &  860.4 & 0.0 & \\  
($\nu$-free) & 1 & 2.06   & 10.07 & 1.18  & 2.74 &  776.3 & 96.2 & 30.4 \\  
    & 2 &   2.11   & 10.66 & 1.08  & 2.51 &  666.3 & 162.9 & 61.9 \\  
       &  4 &   2.37   & 14.60 & 0.70  & 1.54 &  279.8 & 214.0 & 122.6\\  
SLy4  & 1 &   2.16   & 11.00 & 1.13  & 2.51 &  717.7 & 56.2 & 19.5 \\ 
($\nu$-trapped) & 2 &   2.16   & 11.16 & 1.10  & 2.44 &  661.9 & 92.1 & 41.1\\  
       &  4 &   2.14   & 12.07 & 1.03  & 2.30 &  564.9 & 148.9 & 81.9 \\ \hline  
M3Y-P7  & 0 &  2.07  & 10.05 & 1.17  & 2.82 &  869.9 & 0.0 & \\  
($\nu$-free) & 1 & 2.14 & 10.23 & 1.13  & 2.66 &  827.6 & 60.9 & 23.5 \\  
      & 2 &   2.23   & 11.09 & 1.00  & 2.00 &  645.8 & 104.9 & 48.09 \\  
      & 4 &  2.40   & 13.86 & 0.75  & 1.73 &  365.3 & 174.8 &109.5  \\  
M3Y-P7 & 1 &  2.15   & 10.49 & 1.14  & 2.73 &  999.7 & 42.7 &16.1 \\  
($\nu$-trapped) & 2 &   2.22   & 11.09 & 1.07  & 2.44 &  738.9 & 75.7 & 38.0\\  
       &  4 &  2.27   & 11.68 & 1.05  & 2.30 &  693.0 &  138.8 & 77.2 \\ \hline  
D1N    & 0 &  1.23  & 7.75 & 2.36  & 5.24 &  819.9 & 0.0 & \\  
($\nu$-free) & 1 & 1.32 & 8.46 & 1.88  & 4.10 &  581.3 & 97.9 & 22.8\\  
           & 2 & 1.61 & 10.31& 1.29  & 2.82 &  389.3 & 139.9 & 43.1\\  
           & 4 & 2.22 & 14.28& 0.77  & 1.67 &  282.9 & 148.9 & 84.7\\  
D1N        & 1 &  1.82   & 12.20 & 1.07  & 2.05 &  249.5 & 32.8 & 15.1\\  
($\nu$-trapped) & 2 &  1.96   & 12.22 & 1.06  & 2.11 &  314.1 & 64.0 & 31.0\\  
             & 4 &  2.28   & 14.28 & 0.85  & 1.77 &  352.6 & 103.0 & 66.5\\ \hline  
\end{tabular} 
\end{table}

As was done at zero temperature \cite{Loa11}, in order to explore explicitly 
the effects caused by the nuclear symmetry energy to the stable configuration of PNS, 
we have considered solutions of the TOV equations given by the EOS's based on the 
CDM3Y3 and CDM3Y6 interactions \cite{Loa15} and their soft versions CDM3Y3s and 
CDM3Y6s \cite{Loa11}. The only difference between these two groups of the 
CDM3Yn interaction is the modeling of the IV density dependence that gives 
different slopes of the free symmetry energy at high NM densities and 
different temperatures as shown in Fig.~\ref{f4}. One can see in Tables~\ref{t2} and 
Figs.~\ref{f18} and \ref{f19} that in the $\nu$-free case, the impact of the 
symmetry energy is quite strong at entropy per baryon $S/A=0,1$ and 2. For example, 
the PNS maximum gravitational mass obtained with the CDM3Y3s interaction at 
$S/A=1$ is $M_{\rm max}({\rm CDM3Y3s})\approx 1.18~M_\odot$, and it increases to 
$M_{\rm max}({\rm CDM3Y3})\approx 1.64~M_\odot$ when the density dependence of the 
symmetry energy is changed from the soft to the stiff behavior. Similarly with the
CDM3Y6 interaction, we found $M_{\rm max}({\rm CDM3Y6s})\approx 1.65~M_\odot$ and  
$M_{\rm max}({\rm CDM3Y6})\approx 1.97~M_\odot$ at $S/A=1$ (see Fig.~\ref{f18}). 
Thus, the difference in the slope of the symmetry energy at high NM densities could
lead to a difference of $0.3-0.5~M_\odot$ in the predicted $M_{\rm max}$ value.   
We note further that $M_{\rm max}$ obtained with the CDM3Y6 interaction 
at $S/A=0,1$ is quite close to the neutron star mass of $1.97~M_\odot$ observed 
recently by Demores {\it et al.} \cite{Dem10}. Because the isospin dependences 
of the CDM3Y3 and CDM3Y6 interactions are nearly the same, the difference 
of about $0.3~M_\odot$ in the $M_{\rm max}$ values at $S/A=0,1$ obtained with 
these two interactions in the $\nu$-free case should be due to different 
nuclear incompressibilities $K_0$ (see Table~\ref{t1}). Concerning other NN 
interactions under study, only the Sly4 and M3Y-P7 interactions give the 
$M_{\rm max}$ values at $S/A=0,1$ close to the observed neutron star mass 
of $1.97~M_\odot$ \cite{Dem10}. The D1N interaction gives a too low 
$M_{\rm max}\approx 1.23$ and 1.32 $M_\odot$ at $S/A=0$ and 1, respectively, 
while the M3Y-P5 interaction gives unstable results for the PNS configuration
at high temperatures and baryon densities, and are not included in the present
discussion. We note that the instability of the results given by the M3Y-P5 
interaction is purely numerical one that is likely due to the parametrization 
of this version of M3Y-Pn interaction.   

In the presence of trapped neutrinos, the effect caused by different slopes 
of the symmetry energy is diminished already at entropy $S/A=1$ and 2, and  
stable values of the maximum mass and radius of PNS are rather close, independent 
of the behavior of the symmetry energy (see Tables~\ref{t2}, Figs.~\ref{f18} and 
\ref{f19}). We note that the microscopic BHF calculation of the $\nu$-trapped, 
isentropic PNS \cite{Bur10} gives the maximum mass $M_{\rm max}\approx 1.95~M_\odot$ 
at entropy per baryon $S/A=1$ and 2, and the corresponding $R_{\rm max}\approx 10.3$ 
and 10.8 km. Thus, only the PNS configuration obtained with the CDM3Y6 interaction 
has the maximum mass close to the BHF result, while the maximum radius is slightly 
larger than that predicted by the BHF calculation.    

At much higher temperatures of the PNS matter at $S/A=4$, the impact of the symmetry 
energy becomes weaker and similar $\beta$-stable configurations were found with both 
the asy-stiff and asy-soft interactions in both the $\nu$-free and $\nu$-trapped 
scenarios. However, the effect of the increasing entropy on the maximum gravitational 
mass and radius of the $\beta$-stable PNS is very significant, and we found a strong 
increase of $M_{\rm max}$ and $R_{\rm max}$ at entropy per baryon $S/A=4$ compared 
to those obtained at $S/A=0$ (see Tables~\ref{t2} and \ref{t3}, and 
Figs.~\ref{f18}-\ref{f20}). The hot PNS at $S/A=4$ is substantially expanded in its 
mass ($\Delta M_{\rm max}\approx 0.3-0.5~M_\odot$) and size 
($\Delta R_{\rm max}\approx 4-5$ km), while the pressure in the core is decreased 
by a factor up to three. The case of the D1N interaction is extreme with the 
differences $\Delta M_{\rm max}\approx 1~M_\odot$ and $\Delta R_{\rm max}\approx 6.5$ 
km found between the cold neutron star and hot PNS at $S/A=4$ (both being in 
$\beta$-equilibrium without trapped neutrinos). 

With a minor impact of the symmetry energy at entropy $S/A=4$, a very substantial 
difference in temperature of hot PNS matter given by different density dependent 
interactions (shown in the last two columns of Tables~\ref{t2} and \ref{t3}) is 
directly related to the difference in nucleon effective mass illustrated in 
Figs.~\ref{f4n}-\ref{f6n}. While the maximum gravitational masses obtained with 
the D1N, M3Y-P7 and Sly4 interactions are only slightly larger than that obtained 
with the CDM3Y6 interaction by $\Delta M_{\rm max}\approx 0.1~M_\odot$, the difference 
in temperature of PNS at entropy $S/A=4$ is very large, up to about 100 MeV. 
The recent hydrodynamic simulation of the gravitational collapse of a massive 
$40~M_\odot$ protoneutron progenitor to black hole \cite{Hem12} has shown that 
the $\beta$-stable matter (with very small neutrino fraction) in the outer core 
of PNS, is reaching $S/A\approx 4$ at baryon density $\rho_s\approx 0.63\times 
10^{15}$ g/cm$^3$ and temperature $T_s\approx 80-90$ MeV, depending on the inputs 
of the EOS of $\beta$-stable PNS matter (see Fig.~16 of Ref.~\cite{Hem12}). For the 
purpose of comparison, we have also deduced temperature of the PNS layer of baryon 
density of around $0.63\times 10^{15}$ g/cm$^3$ from out mean-field results (see the 
last column of Tables~\ref{t2} and \ref{t3}), and found that only the CDM3Yn and D1N 
interactions give temperature $T_s$ in the $\nu$-free scenario comparable with that 
obtained in the hydrodynamic simulation. The temperature $T_c$ in the center of 
$\beta$-stable hydrostatic PNS given by the CDM3Yn interaction is about 90 and 110 MeV 
in the $\nu$-trapped and $\nu$-free scenarios, respectively. The $T_c$ values given 
by the D1N, M3Y-P7 and Sly4 interactions are much too high (see Tables~\ref{t3}), and 
this effect is clearly due to the unrealistic behavior of nucleon effective mass 
predicted by these interactions at high baryon densities (see Figs.~\ref{f4n} and 
\ref{f5n}). From the above discussion, only the asy-stiff CDM3Yn interaction 
seems to reproduce the configuration of hot PNS comparable with that given by the 
hydrodynamic simulation.   

\begin{figure}\hspace*{-1.0cm}
\includegraphics[width=0.8\textwidth] {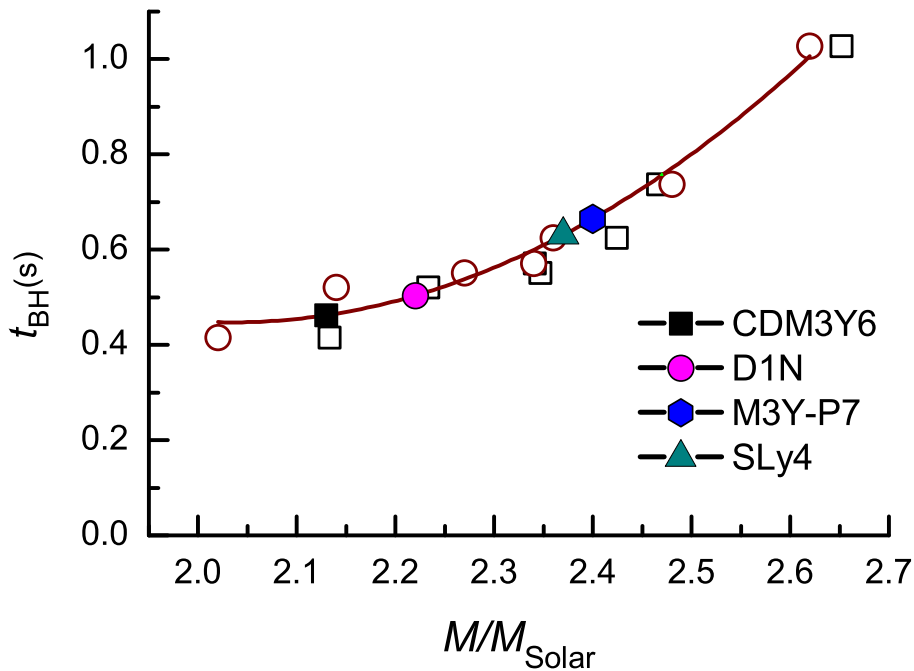} \vspace*{-4cm} 
\caption{(Color online) Delay time $t_{\rm BH}$ from the onset of the collapse of 
a 40~$M_\odot$ progenitor until the black hole formation as function
of the enclosed gravitational mass $M_{\rm G}$ (open squares) given by the 
hydrodynamic simulation \cite{Hem12,Ste13}, and $M_{\rm max}$ values given by the 
solution of the TOV equations using the same EOS for the $\nu$-free and 
$\beta$-stable PNS at $S/A=4$ (open circles). The $M_{\rm max}$ values given by the 
present mean-field calculation of the $\nu$-free and $\beta$-stable PNS at $S/A=4$ 
using different density dependent NN interactions are shown on the correlation 
line interpolated from the results of simulation.} 
\label{f21}
\end{figure}
We discuss our mean-field results for the $\nu$-free and $\beta$-stable PNS 
at $S/A=4$ in more details. Because the simulated density profile of temperature 
reaches its maximum at the total entropy per baryon $S/A\approx 4$, the authors 
of Refs.~\cite{Hem12,Ste13} have compared the stable hydrostatic configurations 
obtained with different EOS's for the $\nu$-free, $\beta$-stable PNS at $S/A=4$ 
with the results of their hydrodynamic simulation. What they found is quite 
remarkable: the gravitational mass $M_{\rm G}$ enclosed inside the shock at the 
onset of collapse of the 40~$M_\odot$ protoneutron progenitor to black hole 
(open squares in Fig.~\ref{f21}) given by the simulation is rather close to the 
maximum gravitational mass $M_{\rm max}$ given by the same EOS of hot PNS matter 
(open circles in Fig.~\ref{f21}). As a result, almost a linear correlation was 
found between the $M_{\rm max}$ value and the delay time $t_{\rm BH}$ from the onset 
of collapse until black hole formation predicted by the simulation \cite{Hem12} 
(see the solid line in Fig.~\ref{f21}). We have combined, therefore, the results 
of this interesting study from Refs.~\cite{Hem12,Ste13}, and presented the correlation 
between $t_{\rm BH}$, $M_{\rm G}$ and $M_{\rm max}$ values in Fig.~\ref{f21}. 
Although the $t_{\rm BH}$ and $M_{\rm G}$ values that might be obtained from 
the hydrodynamic simulation using the EOS's considered in the present work 
are still uncertain, we can use the correlation found between $t_{\rm BH}$ and 
$M_{\rm max}$ in Refs.~\cite{Hem12,Ste13} to roughly estimate $t_{\rm BH}$ based on 
$M_{\rm max}$ values given by the density dependent NN interactions under study. 
Focusing on the CDM3Y6, Sly4, and M3Y-P7 interactions which give the $M_{\rm max}$ 
value at $S/A=0$ close to the observed neutron star mass of around 1.97 $M_\odot$ 
\cite{Dem10}, we find that $t_{\rm BH}\approx 0.5$ s might be obtained with the 
EOS based on the CDM3Y6 interaction, and $t_{\rm BH}\approx 0.6-0.7$ s with the 
EOS's based on the Sly4 and M3Y-P7 interactions. These values seem to satisfy 
the realistic boundary condition $t_{\rm BH}\gtrsim 0.5$ s discussed in 
Ref.~\cite{Hem12}.    

\section{Conclusion}
\label{sec3}
A consistent Hartree-Fock calculation of hot asymmetric NM has been
performed using several realistic choices of the in-medium density dependent NN 
interaction, which have been successfully used in nuclear structure and 
reaction studies. The impact of nuclear symmetry energy and nucleon effective
mass to thermal properties of asymmetric NM has been studied in details. We found 
that the two groups of the interactions (the so-called asy-stiff and asy-soft) that 
give the stiff and soft behaviors of the symmetry energy of NM at zero temperature 
also give the same behaviors of the free symmetry energy of hot NM. The free 
symmetry energy and nucleon effective mass obtained with the asy-stiff CDM3Y3 and 
CDM3Y6 interactions are in a good agreement with the prediction of the microscopic 
BHF calculations \cite{Bur10,Bal14} using the Argonne NN interaction, while those 
results given by the asy-soft interactions differ significantly from the BHF 
results at high NM densities.

We have tested the quadratic approximation for the dependence of the free 
symmetry energy on the neutron-proton asymmetry $\delta$ by comparing 
the results of the HF calculation at different asymmetries $\delta$. While this 
approximation remains reasonable for the internal symmetry energy $E_{\rm sym}/A$ 
at different temperatures, it becomes much poorer for the free symmetry
energy $F_{\rm sym}/A$ with the increasing temperature. Such a breakdown of the 
quadratic approximation is due to the contribution of entropy. The HF calculation
of hot NM has been done in a fine grid of the baryon number densities, neutron-proton 
asymmetries, temperatures and entropies to investigate the density profiles of 
temperature and entropy of hot NM. A very significant impact of the nucleon effective 
mass $m^*$ to the thermodynamic properties of hot NM was found, which is directly related 
to the momentum dependence of the nucleon mean-field potential. At the given baryon
density, the smaller the nucleon effective mass the larger the temperature of NM. 

The density dependent NN interactions were further used to generate the EOS of baryon 
matter in the uniform PNS core, which was smoothly connected to the EOS of the 
inhomogeneous PNS crust by Shen {\it et al.} \cite{Shen,Shen11} for the mean-field
study of $\beta$-stable PNS matter at finite temperature. 
Our study considered two different scenarios: the $\nu$-trapped matter with the 
electron lepton fraction $Y_e\approx 0.4$ and the total entropy per baryon 
$S/A=1,2$ and 4, which mimics the initial stage of PNS; the $\nu$-free matter 
at $S/A=1,2$ and 4, which is close to the late stage of PNS when most neutrinos 
have escaped. The impact of nuclear symmetry energy was found significant in the 
$\nu$-free PNS matter where the dynamic neutron-proton asymmetry, baryon and lepton 
compositions obtained with the asy-stiff and asy-soft interactions are quite different. 
The high-density behaviors of the density profiles of temperature and entropy 
of the $\nu$-free, $\beta$-stable PNS matter are strongly affected not only
by the symmetry energy but also by the nucleon effective mass. Although the impact of 
the symmetry energy was found less significant in the presence of trapped neutrinos, 
we found that the symmetry energy still affects strongly the neutrino fraction 
in the $\nu$-trapped PNS matter, with $x_\nu\lesssim 0.1$ given by the asy-stiff 
interactions at high baryon densities, and $x_\nu\gtrsim 0.1$ given by the asy-soft 
interactions.  
  
Using the inputs of the TOV equation based on different EOS's, we obtained the 
$\beta$-stable hydrostatic configuration of PNS at the total entropy per baryon 
$S/A=1,2$ and 4 in both the $\nu$-free and $\nu$-trapped scenarios. In the absence 
of trapped neutrinos, different slopes of the symmetry energy at high baryon 
densities were shown to give a difference of $0.3-0.5~M_\odot$ in the maximum 
gravitational mass $M_{\rm max}$ predicted with the CDM3Yn interactions. 
For the $\nu$-trapped, $\beta$-stable PNS, the effect of the symmetry energy is 
diminished already at the entropy $S/A=1$ and 2, and the stable values of the 
maximum gravitational mass and radius of the PNS are rather close, independent 
of the behavior of the symmetry energy. However, the impact of the nucleon 
effective mass remains drastic in both $\nu$-trapped and $\nu$-free cases, with 
temperature of the PNS matter being inversely proportional to $m^*$. In particular, 
the difference in temperature in both the outer core and center of PNS given by 
different density dependent NN interactions at $S/A=4$ is mainly due to the difference 
in $m^*$ because the effect of the symmetry energy is diminished at this high entropy  
  
A special attention was given to the configuration of the $\nu$-free PNS at  
entropy $S/A=4$, which was shown by Hempel {\it et al.} \cite{Hem12} to occur at 
the onset of the collapse of a massive (40 $M_\odot$) protoneutron progenitor 
to black hole. We found that at very high temperatures of PNS matter 
at $S/A=4$, the impact of the symmetry energy becomes weaker and the similar 
$\beta$-stable configurations of PNS were obtained with both the asy-stiff and 
asy-soft interactions. The $M_{\rm max}$ and $R_{\rm max}$ values were found 
strongly increased at the entropy per baryon $S/A=4$, with the difference 
$\Delta M_{\rm max}\approx 0.3-0.5~M_\odot$ and $\Delta R_{\rm max}\approx 4-5$ km 
compared to the results obtained at $S/A=0$. Thus, the hot PNS at $S/A=4$ is 
substantially expanded in size, with the decreased central pressure and density. 
In the outer core of hot PNS being at entropy $S/A=4$, only temperature given 
by the asy-stiff CDM3Yn interaction is comparable to that predicted by the 
hydrodynamic simulation. The maximum gravitational masses $M_{\rm max}$ obtained 
for the $\beta$-stable and $\nu$-free PNS at $S/A=4$  using different EOS's of
hot NM were used to estimate the time $t_{\rm BH}$ of collapse of the 40 $M_\odot$ 
progenitor to black hole, based on a correlation found between $t_{\rm BH}$ 
and $M_{\rm max}$ from the hydrodynamic simulation \cite{Hem12,Ste13}.  

In a more general viewpoint, the present mean-field study illustrates the large 
impact of the nuclear symmetry energy and nucleon effective mass in dense and 
hot baryon matter. This effect becomes more complicated after the inclusion 
of trapped neutrinos which weakens the correlation of the thermodynamic properties 
of PNS with the symmetry energy. The densities and temperatures reached in the core 
of PNS at entropy of $S/A\approx 4$ or higher might imply a phase transition 
to new degrees of freedom as discussed, e.g., in Refs.~\cite{Reddy00,Oertel12}. Thus,
PNS is the most extreme compact object which requires the most advanced knowledge 
in the nuclear and QCD physics. It is, however, very short-lived because it lasts only 
a few minutes before being collapsed to black hole or cooled down to neutron star. Thus, 
the observation of the next supernova explosion in our galaxy will certainly provide 
the new and fascinating information about the hot PNS matter that is extremely 
difficult to obtain in the terrestrial nuclear physics laboratories.  

\section*{Acknowledgments}
The present research has been supported, in part, by the National Foundation for 
Scientific and Technological Development (NAFOSTED Project No. 103.04-2014.76),
SNCNS project ANR-10-BLAN-0503, and New-Compstar COST action MP1304. 
N.H.T. is grateful to the University of Lyon for the PALSE incoming mobility 
fellowship that allowed her short-term stay at Institut de Physique Nucl\'eaire 
de Lyon (IN2P3-CNRS), to accomplish an important part of this research.

\appendix
\section{Explicit HF expressions obtained with the finite-range 
CDM3Yn, M3Y-Pn, and Gogny interactions} 
For the interested readers who might want to reconstruct various nuclear EOS's used
in the present mean-field study, we give in this appendix the explicit presentation
of the density dependent NN interactions used in the HF calculation of NM. In general,  
the (central) NN interaction can be given in terms of the spin- and isospin dependent, 
direct (D) and exchange (EX) components as 
\begin{equation} 
v_{\rm D(EX)}(r)=v_{00}^{\rm D(EX)}(r)+v_{10}^{\rm D(EX)}(r)\bm{\sigma}_1\cdot\bm{\sigma}_2
+ v_{01}^{\rm D(EX)}(r)\bm{\tau}_1\cdot\bm{\tau}_2+v_{11}^{\rm D(EX)}(r)
(\bm{\sigma}_1\cdot\bm{\sigma}_2)(\bm{\tau}_1\cdot\bm{\tau}_2), \label{app2}
\end{equation}
where $\bm{\sigma}$ and $\bm{\tau}$ are the Pauli spin and isospin operators, 
respectively. Except of the Sly4 interaction, the CDM3Yn, M3Y-Pn, and Gogny interactions 
are of \emph{finite-range}. For the spin-saturated NM, only the spin-independent 
($v_{00}$ and $v_{01}$) components of the central NN interaction are needed 
in the HF calculation of NM, because the contributions of the spin-dependent terms 
in Eq.~(\ref{app2}) to the NM energy are averaged out. 

Thus, the density dependent CDM3Yn interaction was used in the present HF calculation 
explicitly as
\begin{equation}
 v_{\rm D(EX)}(n_b,r)=F_0(n_b)v^{\rm D(EX)}_{00}(r) + F_1(n_b)v^{\rm D(EX)}_{01}(r)
 \bm{\tau}_1\cdot\bm{\tau}_2,\ {\rm where}\ r=|\bm{r}_1-\bm{r}_2|. \label{app3}
\end{equation}  
The radial parts $v^{\rm D(EX)}_{00(01)}(r)$ are kept unchanged as determined 
from the M3Y-Paris interaction \cite{An83}, in terms of three Yukawas
\begin{equation}
 v^{\rm D(EX)}_{00(01)}(r)=\sum_{k=1}^3 Y^{\rm D(EX)}_{00(01)}(k)
 {{\exp(-a_k r)}\over{a_k r}}. \label{app4}
\end{equation} 
The density dependence of the CDM3Yn interaction (\ref{app3}) was assumed to 
have the same functional form as that introduced first in Ref.~\cite{Kho97}
\begin{equation}
 F_{0(1)}(n_b)=C_{0(1)}[1+\alpha_{0(1)}\exp(-\beta_{0(1)}n_b)+\gamma_{0(1)}n_b].
\label{app5}
\end{equation}
The parameters of the IS density dependence $F_0(n_b)$ were determined \cite{Kho97} 
to reproduce the saturation properties of symmetric NM and give the nuclear 
incompressibility $K=218$ and 252 MeV with the CDM3Y3 and CDM3Y6 versions, 
respectively. The parameters of the IV density dependence $F_1(n_b)$ were determined 
\cite{Loa15} to consistently reproduce the BHF results for the nucleon OP in asymmetric 
NM \cite{Je77,Lej80} in the HF calculation as well as the measured charge-exchange 
$(p,n)$ and ($^3$He,$t)$ data for the isobar analog states in the folding model 
(coupled-channel) analysis \cite{Kho07,Kho14}. The ranges and strengths of Yukawa 
functions and parameters of the density dependence $F_{0(1)}(n_b)$ are given in 
Tables~\ref{tapp1} and \ref{tapp2}, respectively. For the soft version CDM3Yns 
of the CDM3Yn interaction, the IV density dependence was assumed in the same 
functional form as that of the IS density dependence, with the strength slightly 
scaled, $F_1(n_b)\approx 1.1F_0(n_b)$, to obtain the same symmetry energy 
$\varepsilon_{\rm sym}\approx 32$ MeV at the saturation density $n_0$ in the 
HF calculation \cite{Loa11,Kho96}.

As variance with the CDM3Yn interaction, the density dependence of the M3Y-Pn and 
Gogny interactions are introduced via a \emph{zero-range} density dependent term 
\begin{equation}
 v_{00(01)}^{\rm D(EX)}(n_b,r)=v_{00(01)}^{\rm D(EX)}(r)+d_{00(01)}^{\rm D(EX)}(n_b,r). 
\label{app6}
\end{equation}
The first term of the M3Y-Pn interaction (\ref{app6}) has the same finite-range 
form as that given in Eq.~(\ref{app4}), with the Yukawa ranges and strengths 
given in Table~\ref{tapp1}. The direct and exchange parts of the density dependent 
term $d_{00(01)}(n_b,r)$ of the M3Y-Pn interaction are the same and can be given 
explicitly in terms of single-even (SE) and triplet-even (TE) components of the 
two-nucleon force as  
\begin{eqnarray}
 d_{00}^{\rm D}(n_b,r)&=&d_{00}^{\rm EX}(n_b,r)=\frac{3}{8}
\left(t_{\rm SE}{n_b}+t_{\rm TE}{n_b}^{1/3}\right)\delta(\bm{r}), \nonumber\\
 d_{01}^{\rm D}(n_b,r)&=&d_{01}^{\rm EX}(n_b,r)=\frac{1}{8}
\left(t_{\rm SE}{n_b}-3t_{\rm TE}{n_b}^{1/3}\right)\delta(\bm{r}), \label{app7}
\end{eqnarray}
with $t_{\rm SE}=830$ MeV~fm$^6$ and $t_{\rm TE}=1478$ MeV~fm$^4$ adopted for 
the M3Y-P7 version \cite{Na13}.

\begin{table*}
\caption{Ranges and strengths of Yukawa and Gaussian functions used
in the radial dependence of the M3Y-Paris, M3Y-P7, and D1N interactions.}
 \vspace{0.5cm} \label{tapp1}
\begin{tabular}{|c|c|c|c|c|c|c|c|}\hline
Interaction &$k$ & $a_k$& $Y_{00}^{\rm D}(k)$& $Y_{01}^{\rm D}(k)$& 
$Y_{00}^{\rm EX}(k)$ & $Y_{01}^{\rm EX}(k)$ & Reference \\
 & & (fm$^{-1})$& (MeV)& (MeV)& (MeV) & (MeV) & \\ \hline
M3Y-Paris &1 & 4.0 & 11061.625 & 313.625 & -1524.25 & -4118.0 & \cite{An83} \\
          &2 & 2.5 & -2537.5 & 223.5 &-518.75 & 1054.75 & \\
          &3 & 0.7072 & 0.0 & 0.0 & -7.8474 & 2.6157 & \\ \hline
 M3Y-P7   &1 & 4.0 & 12113.375 & 680.625 & -4520.75 & -2945.75 & \cite{Na13}\\
          &2 & 2.5 & -2467.188 & 219.188 & -589.063 & 1059.063 & \\
          &3 & 0.7072 & 0.0 & 0.0 & -7.8474 & 2.6157 & \\ \hline
   D1N    &$k$ & $b_k$& $G_{00}^{\rm D}(k)$ & $G_{01}^{\rm D}(k)$& 
$G_{00}^{\rm EX}(k)$& $G_{01}^{\rm EX}(k)$ & \cite{Ch08} \\ 
 & & (fm) & (MeV)& (MeV)& (MeV) & (MeV) & \\ \hline
       &1 & 0.8 & -369.673 & 827.628  & -26.290 & -338.175 & \\
       &2 & 1.2 & 14.545   & -128.085 & -32.410 & 77.135 & \\ \hline
\end{tabular}         
\end{table*}
\begin{table*}
\caption{Parameters of the density dependence (\ref{app5}) of the CDM3Yn interaction}
 \vspace{0.5cm} \label{tapp2}
\begin{tabular}{|c|c|c|c|c|c|c|}\hline
Interaction & $i$ & $C_i$& $\alpha_i$& $\beta_i$ & $\gamma_i$ & Reference \\
 & & & & (fm$^3$)& (fm$^3$) & \\ \hline
CDM3Y3    &0 & 0.2985 & 3.4528 & 2.6388 &-1.500 & \cite{Kho97} \\
          &1 & 0.2343 & 7.6514 & 9.7494 & 6.6317 & \cite{Loa15} \\ \hline
CDM3Y6    &0 & 0.2658 & 3.8033 & 1.4099 &-4.000 & \cite{Kho97}\\
          &1 & 0.2313 & 7.6800 & 9.6498 & 6.7202 & \cite{Loa15} \\ \hline
\end{tabular}         
\end{table*}
The finite-range part of Gogny interaction is similar to that of the M3Y-Pn 
interaction, but with the radial dependence parametrized in terms of two Gaussians 
\begin{equation}
 v^{\rm D(EX)}_{00(01)}(r)=\sum _{k=1}^{2}G^{\rm D(EX)}_{00(01)}(k) 
 \exp \left(-\frac{r^2}{b^2_k}\right), \label{app8}
\end{equation}
where Gaussian ranges and strengths used with the D1N version are given 
in Table~\ref{tapp1}. The direct and exchange parts of the density dependent 
term $d_{00(01)}(n_b,r)$ of Gogny interaction can be explicitly given as
\begin{eqnarray}
d_{00}^{\rm D}(n_b,r)&=&\frac{t_0}{2}(2+x_0)n_b^{1/3}\delta(\bm{r}), 
\  d_{01}^{\rm D}(n_b,r)=0; \nonumber\\
d_{00}^{\rm EX}(n_b,r)&=&d_{01}^{\rm EX}(n_b,r)=-\frac{t_0}{4}(1+2x_0)
n_b^{1/3}\delta(\bm{r}),  \label{app9}
\end{eqnarray}
with $t_0=1609.46$ MeV~fm$^4$ and $x_0=1$ adopted for the D1N version \cite{Ch08}.

Dividing the total energy density (\ref{ek1}) of NM over the corresponding
baryon number density $n_b$, one obtains the internal NM energy per particle with
the CDM3Yn, M3Y-Pn, and Gogny interactions as 
\begin{equation}
\frac{E}{A}(T,n_b,\delta)=\mathcal{E}_{\rm kin}(T,n_b,\delta)
+\mathcal{E}_{\rm pot}(T,n_b,\delta), \label{ab1}\\
\end{equation}
\begin{eqnarray}
{\rm where}\ \mathcal{E}_{\rm kin}(T,n_b,\delta)&=&\frac{2}{(2\pi)^3n_b}
\sum_{\tau=n,p}\frac{\hbar^2}{2m_\tau}\int n_{\tau}(\bm{k},T) k^2 d{\bm k}; \nonumber\\
 \mathcal{E}_{\rm pot}(T,n_b,\delta)&=&\mathcal{E}_{\rm IS}(T,n_b,\delta)
 + \mathcal{E}_{\rm IV}(T,n_b,\delta), \nonumber\\
 \mathcal{E}_{\rm IS(IV)}(T,n_b,\delta)&=& \mathcal{E}_{\rm IS(IV)}^{\rm D}(n_b,\delta)
+\mathcal{E}_{\rm IS(IV)}^{\rm EX}(T,n_b,\delta).  \label{apb2}
\end{eqnarray}
It is noteworthy that at finite temperature the kinetic energy term is affected 
by different EOS's via the single-particle potential embedded in the nucleon 
momentum distribution $n_{\tau}(\bm{k},T)$. The \emph{direct} term of the (isoscalar 
and isovector) potential energy is obtained compactly with the CDM3Yn interaction as
\begin{equation}
\mathcal{E}_{\rm IS}^{\rm D}(n_b)=\frac{n_b}{2} F_{0}(n_b) J^D_{0},\ \ 
\mathcal{E}_{\rm IV}^{\rm D}(n_b)=\delta^2\frac{n_b}{2}F_{1}(n_b) J^D_{1}, \label{apb3}
\end{equation}
where $J^{\rm D}_{0(1)}=\displaystyle\int v^{\rm D}_{00(01)}(r)d{\bm r}$.
The corresponding \emph{exchange} term can be obtained with the CDM3Yn interaction 
in the following integral form
\begin{eqnarray}
 \mathcal{E}^{\rm EX}_{\rm IS}(T,n_b,\delta)&=&\frac{F_0(n_b)}{8n_b\pi^5}
 \int n_b(\bm k,T)H_{\rm IS}(n_b,{\bm k},\delta,T)d{\bm k}, \nonumber\\
 \mathcal{E}^{\rm EX}_{\rm IV}(T,n_b,\delta)&=&\frac{F_1(n_b)}{8n_b\pi^5} 
 \int\Delta n_b(\bm k,T) H_{\rm IV}(n_b,{\bm k},\delta,T)d{\bm k}. \label{apb4}   
\end{eqnarray}
Here $n_b(\bm k,T)=n_n(\bm k,T)+n_p(\bm k,T),\ \Delta n_b(\bm k,T)
 =n_n(\bm k,T)-n_p(\bm k,T)$, and
\begin{eqnarray}
H_{\rm IS}(n_b,{\bm k},\delta,T)&=&\int n_b(\bm k',T)d{\bm k'} 
\int_0^\infty j_0(kr)j_0(k'r)v^{\rm EX}_{00}(r)r^2dr, \nonumber\\
H_{\rm IV}(n_b,{\bm k},\delta,T)&=&\int \Delta n_b(\bm k',T)
d{\bm k'}\int_0^\infty j_0(kr)j_0(k'r)v^{\rm EX}_{01}(r)r^2dr. \label{apb5}
\end{eqnarray}
Note that the direct term of the potential energy is the same as that obtained at
zero temperature \cite{Loa11,Kho96}, and the explicit temperature dependence is
embedded entirely in the exchange (Fock) term. Thus, a proper treatment of Pauli 
blocking in the mean-field studies of hot NM is indeed very important.  

The exchange term of the potential energy given by the finite-range,  
density independent part of the M3Y-Pn and Gogny interactions can be obtained 
in the same expression (\ref{apb4}), but with $F_{0(1)}(n_b)=1$. It is convenient 
to include the contribution from the zero-range, density dependent part of these 
interactions into the direct term of the potential energy 
\begin{equation}
\mathcal{E}_{\rm IS}^{\rm D}(n_b)=\frac{n_b}{2} \left[J^D_0+D_{\rm IS}(n_b)\right],\  
\mathcal{E}_{\rm IV}^{\rm D}(n_b)=\delta^2 \frac{n_b}{2}\left[J^D_1+D_{\rm IV}(n_b)\right],
\label{apb6}
\end{equation}
where $D_{\rm IS(IV)}(n_b)$ are given explicitly by the M3Y-Pn interaction as
\begin{equation}
 D_{\rm IS}(n_b)=\frac{3}{4}(t_{\rm SE}n_b+t_{\rm TE}n_b^{1/3}),\
 D_{\rm IV}(n_b)=\frac{1}{4}(t_{\rm SE}n_b-3t_{\rm TE}n_b^{1/3}),
\end{equation}
and by Gogny interaction as 
\begin{equation}
 D_{\rm IS}(n_b)=\frac{3}{4}t_0 n_b^{1/3},\ 
 D_{\rm IV}(n_b)=-\frac{t_0}{4}(1+2x_0)n_b^{1/3}.
\end{equation}

The momentum- and density dependent single-particle potential (\ref{uk1})-(\ref{uk2}) 
can be expressed in terms of the IS and IV parts as
\begin{equation}
U_\tau(n_b,\bm k,\delta,T)=U^{(\rm HF)}_{\rm IS}(n_b,\bm k,\delta,T) 
\pm U^{(\rm HF)}_{\rm IV} (n_b,\bm k,\delta,T)+ U^{(\rm RT)}_{\rm IS}(n_b,\delta,T)
 + U^{(\rm RT)}_{\rm IV} (n_b,\delta,T) \label{apu1}
\end{equation}
where the sign + pertains to neutron and - to proton, and
\begin{equation}
 U^{(\rm HF)}_{\rm IS(IV)}(n_b,\bm k,\delta,T)=U^{\rm D}_{\rm IS(IV)}(n_b,\delta)
+U^{\rm EX}_{\rm IS(IV)}(n_b,{\bm k},\delta,T) 
\end{equation}
The \emph{direct} term of the (IS and IV) HF potential is given by the CDM3Yn 
interaction as
\begin{equation}
U^{\rm D}_{\rm IS}(n_b)=n_b F_0(n_b)J^{\rm D}_{0},\ 
U^{\rm D}_{\rm IV}(n_b,\delta)=n_b F_1(n_b)J^{\rm D}_1\delta. \label{apu2}
\end{equation}
The corresponding \emph{exchange} exchange term of the HF potential is given 
by the CDM3Yn interaction as
\begin{equation}
U^{\rm EX}_{\rm IS(IV)}(n_b,{\bm k},\delta,T)=\frac{F_{0(1)}(n_b)}{\pi^2}
H_{\rm IS(IV)}(n_b,{\bm k},\delta,T). \label{apu3}
\end{equation}
The rearrangement term of the single-particle potential (\ref{apu1}) is 
obtained in terms of its direct and exchange terms, and one has in case 
of the CDM3Yn interaction  
\begin{eqnarray}
U^{(\rm RT)}_{\rm IS}(n_b,\delta,T)&=&\frac{\partial F_{0}(n_b)}{\partial n_b}
\left[\frac{n_b^2}{2}J^{\rm D}_0+\frac{1}{8\pi^5} \int n_b(\bm{k},T) 
 H_{\rm IS}(n_b,{\bm k},\delta,T)d{\bm k}\right], \nonumber \\ 
U^{(\rm RT)}_{\rm IV}(n_b,\delta,T)&=&\frac{\partial F_{1}(n_b)}{\partial n_b}
\left[\delta^2 \frac{n_b^2}{2}J^{\rm D}_1+\frac{1}{8\pi^5} \int \Delta n_b(\bm{k},T) 
 H_{\rm IV}(n_b,{\bm k},\delta,T)d{\bm k}\right]. \label{apu4}
\end{eqnarray}
One can see that the explicit temperature dependence of single-particle
potential is entirely embedded in the exchange terms of $U^{(\rm HF)}$ and
$U^{(\rm RT)}$. Moreover, the explicit momentum dependence of single-particle
potential (needed for the estimation of the nucleon effective mass) is also 
contained in the exchange term of the HF potential (\ref{apu3}) only. All this
indicates the vital role of Pauli blocking in dense NM.    

The exchange term of the HF potential given by the finite-range, density 
independent part of the M3Y-Pn and Gogny interactions can be obtained 
in the same expression (\ref{apu3}), but with $F_{0(1)}(n_b)=1$. It is also 
convenient to include the contribution from the zero-range (density dependent) 
part of these interactions into the direct part of the HF potential as
\begin{equation}
U^{\rm D}_{\rm IS}(n_b)= n_b \left[ J^{D}_0 +D_{\rm IS}(n_b) \right], \ 
U^{\rm D}_{\rm IV}(n_b,\delta)= n_b\delta\left[ J^{D}_1 +D_{\rm IV}(n_b)\right].
 \label{apu5}
\end{equation}
As variance with the CDM3Yn case, the rearrangement term of single particle 
potential given by the M3Y-Pn interaction does not depend on temperature
\begin{equation}
U_{\rm IS}^{(\rm RT)}(n_b)=\frac{3}{8}(t_{\rm SE}n_b^2+\frac{1}{3}
t_{\rm TE}n_b^{4/3}),\  U_{\rm IV}^{(\rm RT)}(n_b,\delta)=\frac{\delta^2}{8}
(t_{\rm SE}n_b^2-t_{\rm TE}n_b^{4/3}). \label{apu6} 
\end{equation} 
Likewise, the (temperature independent) rearrangement term of single 
particle potential is given by Gogny interaction as 
\begin{equation}
U_{\rm IS}^{(\rm RT)}(n_b)=\frac{1}{8}t_0 n_b^{4/3},
U_{\rm IV}^{(\rm RT)}(n_b,\delta)=-\delta^2 \frac{t_0}{24}(1+2x_0) n_b^{4/3}. 
\label{apu6} 
\end{equation}

The density- and temperature dependent nucleon effective mass (\ref{eff1}) can be 
explicitly obtained with the CDM3Yn interaction using Eq.~(\ref{apu3}) 
\begin{eqnarray}
\frac{m^*_\tau(n_b,\delta,T)}{m} &=& \left\{1-\frac{m}{\hbar^2 \pi^2 k^\tau_F} 
\left[ F_0(n_b) \int_0^\infty j_1(kr) v_{00}^{\rm EX}(r) r^3 dr 
\int n_b(\bm k',T)j_0(k'r)d {\bm k'} \right.\right. \nonumber\\
&\pm & \left.\left. F_1(n_b) \int_0^\infty j_1(kr) v_{01}^{\rm EX}(r)
r^3 dr \int \Delta n_b(\bm k',T) j_0(k'r)d{\bm k'}\right]\right\}^{-1}, \label{apu7}
\end{eqnarray}
where $j_1(x)$ is the first-order spherical Bessel function, and the + sign pertains 
to neutron and - to proton. The nucleon effective mass in hot NM given by the M3Y-Pn 
and Gogny interactions can be obtained in the same expression (\ref{apu7}), but with 
$F_{0(1)}(n_b)=1$. We emphasize again that the nuclear medium modification responsible
for the nucleon effective mass is fully given by the exchange (Fock) term of single-particle potential.  

\section{Explicit HF expressions obtained with the zero-range Skyrme interaction}
The central zero-range Skyrme interaction \cite{Ch97,Ch98} has been used in the 
present HF calculation explicitly as 
\begin{eqnarray}
v({\bm r})&=& t_0(1+x_0 P_\sigma)\delta(\bm r)+\frac{t_1}{2}
(1+x_1 P_\sigma)[\overleftarrow{\bm P}^2\delta(\bm r)+\delta(\bm r) 
\overrightarrow{\bm P}^2]  \nonumber\\
&+& t_2(1+x_2 P_\sigma)(\overleftarrow{\bm P}.\delta(\bm r)\overrightarrow{\bm P}) 
+ \frac{t_3}{6}(1+x_3 P_\sigma)[n_b(\bm R)]^\gamma\delta(\bm r). \label{sk1} 
\end{eqnarray}
Here ${\bm r}={\bm r}_1-{\bm r}_2,\ {\bm R}=\frac{1}{2}({\bm r}_1+{\bm r}_2),
\ \overrightarrow{\bm P}=\displaystyle\frac{1}{2i}(\overrightarrow{\nabla}_1-
\overrightarrow{\nabla}_2)$, and $\overleftarrow{\bm P}=
\displaystyle\frac{1}{2i}(\overleftarrow{\nabla}_1-\overleftarrow{\nabla}_2)$. 
The spin exchange operator in Eq.~(\ref{sk1}) is taken as 
$P_\sigma=-P_\tau P_r$, where $P_\tau$ and $P_r$ apply the exchange of isospins
(${\bm\tau}_1\rightleftarrows{\bm\tau}_2$) and spacial coordinates 
(${\bm r}_1\rightleftarrows{\bm r}_2$), respectively.

Instead of (\ref{ab1}), the internal NM energy per particle is obtained explicitly
with Skyrme interaction as
\begin{eqnarray}
\frac{E}{A}(T,n_n,n_p)&=&\mathcal{E}_{\rm kin}(T,n_n,n_p)
+\frac{t_0}{4n_b}[(2+x_0)n_b^2-(2x_0+1)(n^2_p+n^2_n)] \nonumber\\
& & + \frac{t_3}{24}n_b^{\gamma-1}\left[(2+x_3)n_b^2-(2x_3+1)(n^2_p+n^2_n)\right],  
\label{sk2} \\
{\rm where}\ \mathcal{E}_{\rm kin}(T,n_n,n_p)&=&\frac{1}{n_b}\sum_{\tau=n,p}
\frac{\hbar^2\mathcal{K}_\tau(T)}{2m^*_\tau},\ {\rm with}\ 
\mathcal{K}_\tau(T)=\frac{2}{(2\pi)^3}\int n_{\tau}(\bm{k},T)k^2 d{\bm k}. \label{sk3}
\end{eqnarray}
Due to the zero range of Skyrme interaction, the temperature dependence implied 
by the nucleon momentum distribution is integrated out and the potential energy 
term in Eq.~(\ref{sk2}) turns out to be the same as that at zero temperature. 
As a result, the temperature dependence of the internal NM energy is entirely 
embedded in the kinetic energy term (\ref{sk3}), where the (temperature 
independent) nucleon effective mass $m^*_\tau$ is determined as 
\begin{eqnarray} 
\frac{m^*_\tau(n_n,n_p)}{m}&=&\left[1+\frac{m}{4\hbar^2}
(\Theta_1n_b+\Theta_2n_\tau)\right]^{-1}, \label{sk4} \\
{\rm with}\ \ \Theta_1 &=& t_1(2+x_1)+t_2(2+x_2), \nonumber\\
           \Theta_2 &=& t_2(2x_2+1)-t_1(2x_1+1). \nonumber
\end{eqnarray}
The momentum- and density dependent single-particle potential (\ref{uk1})-(\ref{uk2}) 
can be obtained with Skymre interaction explicitly as
\begin{eqnarray}
U_\tau(n_n,n_p,{\bm k},T)&=&U^{(\rm HF)}_\tau(n_n,n_p,{\bm k},T)+
U^{(\rm RT)}(n_n,n_p), \nonumber\\
U^{\rm (HF)}_\tau(n_n,n_p,{\bm k},T)&=&\frac{t_0}{2}[(2+x_0)n_b-(2x_0+1)n_\tau] 
+\Theta_1[{\mathcal{K}_n(T)+\mathcal{K}_p}(T)]+\Theta_2\mathcal{K}_\tau(T)\nonumber\\
& & +\frac{t_3}{12}n_b^{\gamma}\left[(2+x_3)n_b-(2x_3+1)n_\tau \right]
+ \frac{1}{8}\left[\Theta_1 n_b + \Theta_2 n_\tau \right]{\bm k}^2, \label{sk5}\\
{\rm and}\ U^{\rm (RT)}(n_n,n_p)&=& \frac{t_3}{24}\gamma n_b^{\gamma-1}
\left[(2+x_3)n_b^2-(2x_3+1)(n_n^2+n_p^2)\right]. \label{sk6}
\end{eqnarray}
Like the potential energy term in Eq.~(\ref{sk2}), the temperature dependence of 
single particle potential is also given by that of the kinetic energy term, via 
$\mathcal{K}_\tau(T)$ values. With the explicit momentum dependence of Skyrme 
interaction (\ref{sk1}), the HF term of single-particle potential (\ref{sk5}) has
a quadratic momentum dependence that gives the nucleon effective mass (\ref{sk4}). 
Like the case of M3Y-Pn and Gogny interactions, the rearrangement term (\ref{sk6}) 
of Skyrme single-particle potential is also temperature independent.  

The explicit parameters of the SLy4 version \cite{Ch98} of Skyrme interaction used
in the present HF calculation are: $t_0$=-2488.91 MeV~fm$^3$, 
$t_1$=486.82 MeV~fm$^5$,  $t_2$=-546.39 MeV~fm$^5$, $t_3$=13777.0 MeV~fm$^{3+3\gamma}$,
$x_0$=0.834, $x_1$=-0.344, $x_2=$-1.000, $x_3$=1.354, and $\gamma=1/6$.
\end{document}